\documentclass[preprint2]{aastex} 

\usepackage{natbib}

\begin{document}

\title{Star formation activity and gas stripping in the Cluster
  Projected Phase-Space (CPPS)}

\author{Jonathan D. Hern\'andez-Fern\'andez\altaffilmark{1}
  , 
  C. P. Haines\altaffilmark{2} 
  , 
  A. Diaferio\altaffilmark{3,4}
  , 
  J. Iglesias-P\'aramo\altaffilmark{5,6} 
  , 
  C. Mendes de Oliveira\altaffilmark{1}
  and 
  J. M. Vilchez\altaffilmark{5} 
  }

\altaffiltext{1}{Departamento de Astronomia, Instituto de Astronomia,
  Geof\'isica e Ci\^encias Atmosf\'ericas da Universidade de S\~ao
  Paulo, Rua do Mat\~ao 1226, Cidade Universit\'aria, 05508-090, S\~ao
  Paulo, Brazil; jonatan.fernandez@iag.usp.br}

\altaffiltext{2}{Observatorio Astron\'omico Cerro Cal\'an,
  Departamento de Astronom\'ia, Universidad de Chile, Casilla 36-D
  Santiago, Chile}

\altaffiltext{3}{Dipartimento di Fisica, Universit\`a di Torino,
  V. Pietro Giuria 1, 10125 Torino, Italy}

\altaffiltext{4}{Istituto Nazionale di Fisica Nucleare (INFN), Sezione
  di Torino, V. Pietro Giuria 1, 10125 Torino, Italy}

\altaffiltext{5}{Instituto de Astrof\'isica de Andaluc\'ia, Glorieta
  de la Astronom\'ia s/n, 18008 Granada}

\altaffiltext{6}{Centro Astron\'omico Hispano Alem\'an C/ Jes\'us
  Durb\'an Rem\'on, 2-2 04004 Almer\'ia}

\begin{abstract}

This work is focused on the study of the distribution in the CPPS of
passive(ly-evolving) and star-forming galaxy populations and also, the
intense and quiescent star-forming populations for a set of nine
nearby $z{<}$0.05 galaxy clusters. Furthermore, we compare the CPPS
distribution of the passive galaxy population with the accreted halo
population of a set of 28 simulated clusters and the star-forming
population with the non-accreted population. We consider various
cluster accretion epochs and accretion radii where it is assumed that
star formation in galaxies becomes quenched, in order to segregate the
accreted population from the non-accreted population. Just applying
this segregation in simulations, we get a qualitative agreement
between the CPPS distributions of the passive and the accreted
populations and also between the star-forming and the non-accreted
populations. The uncertainty in cluster centering strongly affects the
pronounced cuspy profiles of the projected density and also, it can
explain the main difference (i.e. inner slope) between the CPPS
distribution of passive and accreted populations. The CPPS density of
star-forming galaxies and the intensity of ram-pressure stripping
present an opposite trend throughout the CPPS. This implies that
ram-pressure stripping significantly contributes to modulate the
observed CPPS distribution of star-forming galaxies in cluster virial
regions and their surroundings. The significant fraction of
star-forming galaxies at the projected center of clusters are mainly
those galaxies with low l-o-s velocities and they can be mainly
identified as those galaxies with a remaining star formation activity
(quiescent star-forming galaxies) inside the physical virial region
or, in a lower degree, as galaxy interlopers i.e. outside the physical
virial region. This article also includes a test of the effects caused
by the Sloan fibre collision on the completeness of the Main Galaxy
Sample as a function of clustercentric radius.


\end{abstract}



\section{Introduction}
\label{sec:intro} 

Galaxies, especially in clusters, are under the influence of a number
of environmental processes producing a strong impact on their stellar
and gaseous components and their global properties as morphology, star
formation activity, etc. The specific contribution of each one of
these processes, which can operate alone or simultaneously, to the
modulation of galaxy properties is an issue which is not totally
clarified so far. Several authors \citep[e.g.][]{Dickens&Moss_1976,
  Colless&Dunn_1996,Mohr_et_al_1996b,Biviano_et_al_1997,
  Carlberg_et_al_1997a,Fisher_et_al_1998,ENACS_VI,ENACS_XI,
  Diaferio_et_al_2001} have pointed out that the combined study of the
spatial and kinematic variables of galaxy distributions in clusters as
a function of galaxy properties and galaxy populations, is key to
shedding light on how the cluster galaxy population is formed. The
most appropriate space which combines these variables is the Cluster
Projected Phase-Space (CPPS). This space is defined by the
clustercentric projected radius $R_{P}$ normalized by the virial
radius $r_{200}$, \mbox{$\tilde{r}$=$R_{P}$/$r_{200}$} and the
cluster-frame line-of-sight velocity c$\Delta{z}$/(1+$z_{c}$)
normalized by the rest-frame cluster velocity dispersion $\sigma_{c}$,
$\tilde{s}{=}$(c$\Delta{z}$/(1+$z_{c}$))/$\sigma_{c}$.

A precise estimation of the impact of environmental processes along
the history of a galaxy orbiting its parent cluster is rather
difficult due to the nature of galaxy clusters. First of all, the
observed velocity dispersions of ${\sim}1000$\,km\,s$^{-1}$ for rich
galaxy clusters imply that we normally cannot measure the l-o-s
distance to a given spectroscopic cluster member to an accuracy better
than ${\sim}15$\,Mpc. Thus, we cannot identify whether it is in the
cluster core, in the outskirts, or even whether it has ever felt any
environmental influence of the cluster. Furthermore, clusters contain
a backsplash galaxy population which is composed of those galaxies
which, after reaching the pericenter of their orbits and suffering the
environmental effects of virial regions, currently show a position
well outside the virial region \citep{Gill_et_al_2005,
  Sato&Martin_2006,Pimbblet_2011,Mahajan_et_al_2011b,Oman_et_al_2013}.
This population has been invoked to explain the presence of
HI-stripped galaxies out to large radii outside the central regions of
clusters \citep{Solanes_et_al_2001}. In addition, a pre-processing
scenario \citep{Fujita_2004} is proposed for those galaxies mainly
affected by tidal interactions within the frame of galaxy groups or
filaments feeding galaxy clusters but outside the virial region
\citep{Porter_et_al_2008,Bahe_et_al_2013,
  Dressler_et_al_2013,Lopes_et_al_2013}, with the case of the Blue
Infalling Group \citep{Iglesias-Paramo_et_al_2002,Cortese_et_al_2006}
as the prototype.

Nevertheless, some properties of galaxy population as star-forming
activity, morphology or galaxy luminosity show a significant degree of
segregation as a function of the environmental conditions. These
environmental conditions are spatial conditions e.g. cluster-centric
radius, volume density of galaxies, or kinematical conditions such as
the velocity distribution of the galaxy system or the type of orbits
around the cluster. Here, we outline a few of these trends. The
H$\alpha$ emission as a tracer of the very recent star-formation
activity presents a trend in fraction of star-forming galaxies and the
intensity of star-formation activity depending on environmental
conditions as the clustercentric radius
\citep[e.g.][]{Lewis_et_al_2002,Balogh_et_al_2004,Koopmann&Kenney_2004,
  CAIRNS_III,Hwang&Lee_2008}. There are a number of works which found
that late-type galaxies in clusters have statistically significant
higher velocity dispersion than that shown by early-type galaxies
\citep[e.g.][]{Sodre_et_al_1989,ENACS_XIII,Goto_2005}. Regarding the
galaxy luminosity, the brightest cluster galaxy (or the set of
brightest galaxies) shows a systematic lower velocity departure from
the cluster average velocity than the rest of the galaxies
\citep[e.g.][]{Adami_et_al_1998,Goto_2005,Hwang&Lee_2008} or in some
cases well away from the mean cluster velocity
\citep{Pimbblet_et_al_2006}. Recently, \citet{Haines_et_al_2012} have
presented a study of the CPPS distribution of X-ray active galactic
nuclei (AGN) for a representative sample of massive clusters at
intermediate redshifts, providing evidence that X-ray AGNs found in
massive clusters are an infalling population. Assuming three different
categories of orbits (radial, circular and isotropic) depending on the
budget between the radial component and the tangential component of
the galaxy velocity vector; some authors find a distinct orbital
behaviour for different (morphological and/or spectral) galaxy
populations \citep[e.g.][]{Ramirez&deSouza_1998,
  ENACS_XIII,ENACS_VII}. In contrast, other works claim there is no
significant difference between orbital distribution for different
cluster galaxy populations
\citep[e.g.][]{CAIRNS_I,Van_der_Marel_et_al_2000,Goto_2005}.

Assuming the CPPS contains two prominent observables (clustercentric
radius and cluster-frame line-of-sight velocity) which strongly
determine the intensity of the different environmental processes, it
is crucial to the study the distribution of the different galaxy
populations in this phase-space, in order to assess the specific
influence of each process in the build up of each galaxy
population. In this article, we focus on the study of the observed
distributions of galaxies in the CPPS depending on their level of
star-formation activity. We will deal with two galaxy dichotomies;
passive vs. star-forming galaxies and quiescent vs. intense
star-forming galaxies. In order to divide cluster galaxies between
passive and star-forming galaxies, we take advantage of vacuum
ultraviolet (UV) data from GALEX (Galaxy Evolution Explorer)
\citep{Morrissey_et_al_2005, Martin_et_al_2005}. The vacuum UV stellar
emission is a robust tracer of the recent star formation because it
comes from the more short-lived stars $\tau{<}$10$^{8}$ yr
\citep[][]{Kennicutt_1998,Kauffmann_et_al_2007, Martin_et_al_2005}.

The remainder of this paper is organised as follows. In section
\ref{sec:gal_samp} we describe the set of galaxy clusters studied in
this work and the improved estimation of their cluster properties. In
section \ref{sec:results}, we present the results on the segregation
of passive and star-forming galaxies in the CPPS in subsection
\ref{ssec:pe_sf_cpps} and the expected distribution of galaxy halos in
the CPPS depending on the accretion epoch and the accretion radius
i.e. the time and the radius where it is assumed the star formation in
a galaxy starts to be quenched, dividing them between formerly
accreted and non-accreted halos in subsection
\ref{ssec:mill_cpps}. The motivation of Section \ref{sec:discussion}
is establishing links between the accretion history of cluster
galaxies and their star-formation mode, passive or
star-forming. Assuming the global shape of a CPPS distribution comes
from the combination of the velocity distribution and the radial
profile of the projected density, we discuss the comparison of these
two observables between the halo populations and the galaxy
populations in detail in this section. In subsection
\ref{ssec:appendixC}, we discuss about the effect of uncertainties of
cluster centering in the derivation of the density profile and the
effect of uncertainties of cluster average redshift in the derivation
of the velocity distribution. In subsection
\ref{ssec:stripp_intensity} we assess and describe the influence of
the ram-pressure on the modulation of the star-formation activity in
the framework of the CPPS and in subsection \ref{ssec:stfm_in_cpd} we
probe and describe the distribution of star-forming galaxies split by
their intensity of star-formation activity. In section
\ref{sec:summ&conclu} we summarize the main results and conclusions of
this work. In Appendix, we compute the effect of incompleteness of the
Main Galaxy Sample \citep{Strauss_et_al_2002} as a function of
clustercentric radius. Throughout the paper, all physical magnitudes
are computed assuming a cosmological model characterized by the
following parameters: $h{=}$0.7, $\Omega_{m}{=}$0.3 and
$\Omega_{\Lambda}{=}$0.7.


\section{The set of galaxy clusters}
\label{sec:gal_samp}

The set of galaxy clusters under study in this work is taken from the
sample of cluster galaxies extensively described in \citet[][hereafter
  Paper I]{Hernandez-Fernandez_et_al_2012a}. This sample consists of a
total of more than 5000 galaxies distributed in 16 nearby clusters
($z{<}$0.05) showing a rich variety in their properties, from poor to
rich clusters.  The galaxy sample spans an $r'$-band luminosity range
corresponding to $-$23${\lesssim}M_{\it\,r'}{\lesssim}{-}$18.  The
faint limit in this magnitude range corresponds approximately to the
classical luminosity boundary between giant and dwarf galaxies
$M_B{\sim}-$18. The selection of this sample was constrained by the
condition that it be covered by the Data Release 6 of the Main Galaxy
Sample (MGS) of SDSS \citep{Adelman-McCarthy_et_al_2008} and the AIS
(All Imaging Survey) of GALEX mission \citep{Martin_et_al_2005}. We
restrict the UV data to the AIS in order to keep a homogeneous
completeness in the UV bands for the galaxy sample. The completeness
limit in AB-magnitude corrected for Galactic extinction of
NUV$\sim$22.5 for the SDSS-GALEX matched catalog
\citep{Bianchi_et_al_2007} and the completeness limit of the SDSS MGS
in the $r'$-band composite-model magnitude $r'{=}$17.56 allow us to
robustly identify and segregate star-forming galaxies from passive
galaxies down to the MGS SDSS magnitude limit \citep[see Figure 4
  of][]{Hernandez-Fernandez_et_al_2012b}.

Given that this work is focused on the study of the CPPS, we require
an accurate estimation of the cluster center, the average redshift and
the velocity dispersion of each cluster. For this purpose, we rely on
the caustic method \citep{Diaferio&Geller_1997,Diaferio_1999}. The
caustic method, originally proposed by \citet{Diaferio&Geller_1997},
is an approach which provides the mass profile and the profile of the
escape velocity along the line of sight $\langle
v^{2}_{\rm\,esc}\rangle_{\rm\,l\cdot\,o\cdot\,s}$ out to cluster
regions well beyond $r_{200}$, using only the galaxy celestial
coordinates and redshifts, and without assuming dynamical equilibrium
for the cluster \citep{Diaferio_1999}. It is important to say that the
caustic method includes the hierarchical clustering method
\citep{Serna&Gerbal_1996,Diaferio_1999} for the cluster center
determination which is unbiased with respect to the center of mass of
the cluster. We note that the velocity dispersion retrieved from a set
of simulated clusters through the caustic method in
\citet{Serra_et_al_2011} is in good agreement with the {\it true}
velocity dispersion, defined as the line-of-sight velocity dispersion
of the particles within a sphere of radius 3$r_{200}$ in real
space. In fact, the caustic velocity dispersion was found to be within
5\,\% of the true value for the 50\,\% of the cluster set and within
30\,\% of the true value for the 95\,\% of the cluster
set. \citet{Serra_et_al_2011} also show that a few tens of redshifts
per squared comoving megaparsec within the cluster, are enough to
derive a reasonably accurate estimate of the escape velocity profile
out to around 4$r_{200}$. Regarding the cluster centering, the caustic
method provides cluster centers which differs, on average,
approximately 150 kpc from the X-ray centroids for a X-ray cluster
catalogue of 72 clusters in \citet{CIRS}. We apply the caustic method
to a set of nine massive ($\sigma_{c}{\gtrsim}$500\,km\,s$^{-1}$)
clusters; the resulting cluster centers, average redshifts, rest-frame
velocity dispersions and other derived cluster properties are reported
in Table \ref{tab:CS}. In this cluster set, we include the seven
massive clusters ($\sigma_{c}{>}$500\,km\,s$^{-1}$) from the original
sample and also ABELL\,2197 and WBL\,518. We include just massive
clusters to avoid the poor statistics and the less accurate cluster
properties (cluster center, velocity dispersion, etc) provided by poor
galaxy groups. The two clusters explicitly outlined are included in
the galaxy sample described in \citet{Hernandez-Fernandez_et_al_2012a}
but initially they were not selected as `central' clusters in the
original sample \citep[see section 3
  in][]{Hernandez-Fernandez_et_al_2012b}.


\begin{table}[!ht]
\caption{\bf Caustic properties of the observed cluster set.}
\begin{center}
\resizebox{1.00\hsize}{!}{
\begin{tabular}{c              c              c              c            }
\hline
Cluster       & $\alpha_{c}$  & $\delta_{c}$ & $z_{c}$      \\
              & deg          & deg          &             \\
 (1)          & (2)          & (3)          & (4)         \\     
\hline        
ABELL 671     &  127.133   &    30.415  &    0.04952      \\  
ABELL 1185    &  167.654   &    28.691  &    0.03325      \\ 
ABELL 1213    &  169.125   &    29.266  &    0.04671      \\ 
WBL 514       &  218.518   &     3.783  &    0.02864      \\ 
WBL 518       &  220.170   &     3.459  &    0.02717      \\ 
UGCl 393      &  244.447   &    35.035  &    0.03160      \\ 
B2(*)         &  245.742   &    37.943  &    0.03077      \\ 
ABELL 2199    &  247.144   &    39.552  &    0.03035      \\ 
ABELL 2197E   &  247.483   &    40.650  &    0.03024      \\ 
\hline
\\
\hline
  $z_{GA}$     &  $\sigma_{c}$  & $r_{200}$     &  $M_{200}$          \\
              & km s$^{-1}$    & Mpc          &   10$^{14}M_{\odot}$  \\ 
(5)           & (6)           &   (7)        &    (8)              \\     
\hline
      0.05013 &   715.16      &    1.727     &   6.127  \\ 
      0.03476 &  1165.24      &    2.834     &  26.694  \\ 
      0.04813 &   682.02      &    1.649     &   5.319  \\ 
      0.03069 &   492.76      &    1.201     &   2.023  \\ 
      0.02923 &   505.55      &    1.233     &   2.186  \\ 
      0.03314 &   505.63      &    1.231     &   2.183  \\ 
      0.03225 &   508.74      &    1.239     &   2.224  \\ 
      0.03180 &   742.21      &    1.808     &   6.908  \\ 
      0.03168 &   692.85      &    1.688     &   5.620  \\
\hline
\end{tabular}}   
\end{center}
{\bf (1) NED Object Name of the cluster. The complete NED name of
  B2(*) is \mbox{B2 1621+38:[MLO2002] CLUSTER}, (2) and (3) Celestial
  coordinates of cluster center, (4) average redshift and (5) redshift
  corrected from nearby attractors computed with Moustakas IDL
  procedure {\it mould\_distance.pro} from
  http://code.google.com/p/idl-moustakas/ following
  \citet{Mould_et_al_2000} (6) rest-frame cluster velocity dispersion
  computed through the caustic method, (7) $r_{200}$ and (8) $M_{200}$
  are, respectively, the virial radius and the virial mass computed
  following \citet[][]{Finn_et_al_2005} using the $\sigma_{c}$ as
  input.}
\label{tab:CS}
\end{table}

The comparison of the {\it caustic} velocity dispersions with respect
to the values of velocity dispersion computed through the
\citet{Poggianti_et_al_2006}'s procedure (see Paper I) produces an
agreement within ${\Delta}{\sigma_{c}}{\lesssim}$150\,km\,s$^{-1}$
among the seven clusters in common with the original sample and of the
same order of magnitude of the uncertainty of the cluster velocity
dispersion ${\delta}{\sigma_{c}}{\sim}$50-100\,km\,s$^{-1}$, except
for the case of ABELL\,1185, the more massive cluster, with a
difference of \mbox{${\Delta}{\sigma_{c}}{\sim}$375 km s$^{-1}$}. In
the case of the offsets in celestial coordinates of cluster centers,
they are less than \mbox{${\Delta}{\theta}{\sim}$3.15 arcmin} between
clusters in common with the original sample. This offset corresponds
to ${\sim}$115 kpc in physical distance at $z$=0.03. We stress that
the {\it caustic} cluster center as a more robust estimation than the
values reported by NED insofar it is unbiased with respect to the
center of mass of the cluster.

\section{Results: Distribution of galaxies in the CPPS}
\label{sec:results}

\subsection{Passive and star-forming cluster galaxies in the CPPS}
\label{ssec:pe_sf_cpps}

In order to separate the star-forming galaxies from the
passively-evolving galaxies, we apply the UV-optical color cut
proposed for the (NUV${-}r'$) vs. ($u'{-}r'$) color-color diagram in
Paper I. Specifically, we assume a galaxy is a star-forming galaxy if
it fulfils one of the following conditions: \\ $u'{-}r'{<}$2.175 and
NUV${-}r'{<}$4.9 \\ $u'{-}r'{>}$2.175 and
NUV${-}r'{<}-$2($u'{-}r'$)+9.25 or \\ $u'{-}r'{<}$2.22 and there is no
GALEX counterpart for this specific galaxy. \\ Otherwise, we classify
the galaxy as a passive galaxy.

The integrated NUV${-}r'$ colour is enough to provide a robust
separation of passive and star-forming galaxies
\citep{Salim_et_al_2005,Kauffmann_et_al_2007,Martin_et_al_2007} with a
scarce presence (less than $\sim$10 per cent) of H$\alpha$
emission-line galaxies in the UV-optical red sequence
\citep{Haines_et_al_2008}. Also, the combined UV-optical color-color
diagrams are suitable to detect galaxies with truncated star formation
histories \citep{Kaviraj_E+A}. Furthermore, the integrated ultraviolet
and optical colors present the advantage with respect to the fiber
spectra of not suffering from the aperture bias effect
\citep[e.g.][]{Kewley_Jansen&Geller_2005}. This effect easily produces
the miss-classification of those early-spiral galaxies which appear
passive from a spectrum that samples only their bulge, but have also
normal star-forming discs \citep{Haines_et_al_2008}.

\begin{figure}[!hp]
\centering
\resizebox{1.00\hsize}{!}{\includegraphics{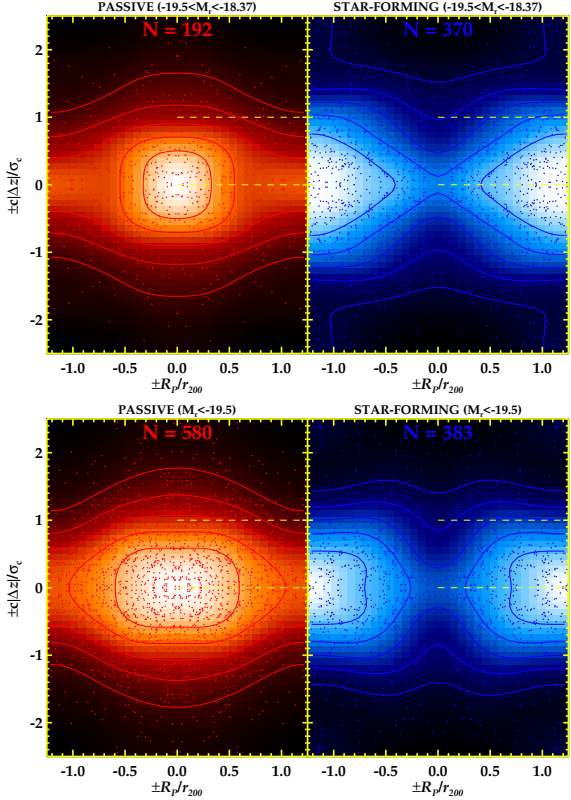}}
\caption{\bf Stacked CPPS of low-luminosity
  ($-$19.5${<}M_{\it\,r'}{\le}-$18.37) (top panel) and high-luminosity
  ($M_{\it\,r'}{\le}-$19.5) (bottom panel) clusters galaxies. The
  diagrams are symmetrized with respect to both axes. Left (right)
  panel corresponds to passive (star-forming) galaxies. The intensity
  map corresponds to the CPPS density of galaxies and contours
  represent the lines of isodensity. Two parallel horizontal dashed
  lines are located at $\tilde{s}$=0 and $\tilde{s}$=1 only for
  comparison purposes. The numbers at the top indicate the numbers of
  galaxies which are in each quadrant of the plot.}
\label{fig:high-lum_gal1}
\end{figure}

\begin{figure}[!ht]
\centering
\resizebox{1.00\hsize}{!}{\includegraphics{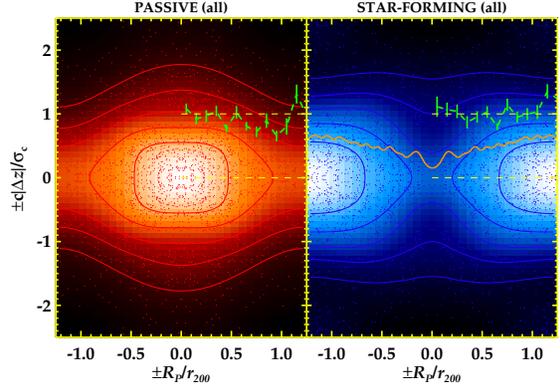}}
\caption{\bf Stacked CPPS of the whole cluster galaxy sample. The
  broken dashed lines with uncertainty bars represent the radial
  profile of velocity dispersion and vertical bars are the bootstrap
  uncertainties in each bin. The solid line in the right panel shows
  the radial trend of the star-forming galaxy fraction (see the
  vertical axis as reference for numerical values). Color code,
  isodensity lines and the rest of elements of the figure are defined
  in the same way as figure \ref{fig:high-lum_gal1}.}
\label{fig:all_gal1}
\end{figure}

In figure \ref{fig:high-lum_gal1} and figure \ref{fig:all_gal1}, we
plot the distribution of cluster galaxies in the CPPS depending on
their star formation activity and luminosity. In figure
\ref{fig:high-lum_gal1}, we show the whole galaxy sample split in two
sub-samples of similar size for magnitudes brighter and fainter than
$M_{\it r'}{=}{-}$19.5 \citep[this corresponds approximately to
  $M^{\ast}_{r'}$+1.70,][]{SDSS_LF}. Figure \ref{fig:all_gal1} shows
the complete cluster galaxy sample down to
$M_{\it\,r'}{=}{-}$18.37. In order to enhance the statistics and
taking advantage of the symmetry of clusters in the kinematic axis
with respect to the average redshift, especially in the virial region,
we merge the positive and negative quadrants of the CPPS in only one
quadrant. We return to discuss the accuracy of this assumption later
in this section, also see table \ref{tab:KS_symm}. In addition, we
mirror this quadrant with respect to both axes with the motivation
that this approach makes the visualization, description and comparison
of CPPS distributions between different galaxy populations, easier.

All density maps shown in this article are 33 bin$\times$55 bin
two-dimensional histograms smoothed with a two-dimensional Gaussian
kernel of $\sigma(\tilde{r}){=}\frac{5}{6}$ and
$\sigma(\tilde{s})$=1. We choose this relatively large size of
Gaussian kernel to derive more robust statistical properties and to
dilute the galaxy substructures observed in galaxy clusters even near
to the projected cluster center \citep[][]{ENACS_XI}. More important,
our aim is studying the global CPPS distribution of clearly distinct
galaxy populations e.g. passive vs. star-forming, more than to give a
precise computation of the small-scale variations of the observed
distributions of galaxy populations. Iso-density contours are equally
spaced and divide the whole range of density for each CPPS
distribution in six bins. With respect to the radius range shown in
previous figures, we check that outside $R_{P}{=}$0.9$r_{200}$ the
contribution of galaxies from neighboring clusters becomes noticeable.


In a first inspection to figures \ref{fig:high-lum_gal1} and
\ref{fig:all_gal1}, the two galaxy populations show clearly distinct
distributions. While the distribution of passive galaxies is clearly
concentrated around the origin of coordinates
\mbox{($\tilde{r}$,$\tilde{s}$)${\sim}$(0,0)}, the distribution of
star-forming galaxies is biased towards radii close to and outside the
projected virial radius. On the contrary, the distribution of the two
bins of different luminosity ranges, $M_{\it\,r'}{\le}-$19.5 and
$-$19.5${<}M_{\it r'}{\le}-$18.37 of the same galaxy population
(passive or star-forming) seem quite similar. In order to make a
quantitative comparison of the CPPS distributions,
$\mathcal{D}$($\tilde{r}$,$\tilde{s}$), for the two luminosity bins of
the same galaxy population, we apply a Kolmogorov-Smirnov (K-S) test
to the CPPS distributions in the CPPS region delimited by
0${<}\tilde{r}{<}$0.9 and 0${<}\tilde{s}{<}$3. The values for the
probabilities and respective uncertainties, that the two distributions
come from the same parent function are quoted in table
\ref{tab:KS_highlowlum}. The uncertainties here are computed assuming
binomial uncertainties for the cumulative distribution functions of
each data distribution. The purpose of including the uncertainties is
showing the effect of the sample size in the accuracy of the K-S
probabilities i.e. two distributions of very different sizes can give
the same K-S probability but these two values should not be equally
significant. Also, we show in table \ref{tab:KS_symm} the results of
different K-S tests applied to CPPS distributions of galaxy
populations in order to check the assumption of symmetry of clusters
in the kinematic axis with respect to the average l-o-s
velocity. These tests are described in the corresponding caption.


\begin{table}[!ht]
\caption{\bf Results from the Kolmogorov-Smirnov test to assess the
  similarity of the CPPS distributions of (passive and star-forming)
  galaxy populations in different luminosity bins:
  $M_{\it\,r'}{\le}-$19.5 (HL) and $-$19.5${<}M_{\it\,r'}{\le}-$18.37 (LL).}
\begin{center}
\resizebox{1.00\hsize}{!}{
\begin{tabular}{l       c  }
\hline
{\sc Galaxy population}   
                    & $\mathcal{D}_{HL}$($\tilde{r}$,$\tilde{s}$) $\sim$ 
                      $\mathcal{D}_{LL}$($\tilde{r}$,$\tilde{s}$) \\
                    & (1)                \\   
\hline
{\sc PASSIVE}       &  69$^{+18}_{-19}$ \%    \\ 
{\sc STAR-FORMING}  &  47$^{+21}_{-17}$ \%    \\ 
\hline
\end{tabular}} 
\end{center}
{\bf (1) K-S probability of ($\tilde{r}$,$\tilde{s}$) 2D distributions
  of high ($M_{\it r'}{\le}-$19.5) and the low-luminosity
  ($-$19.5${<}M_{\it\,r'}{\le}-$18.37) cluster galaxies come from the
  same parent 2D function.}
\label{tab:KS_highlowlum}
\end{table}


\begin{table}[!hb]
\caption{\bf Results for the Kolmogorov-Smirnov test to assess the
  symmetry in the kinematic axis of the CPPS distributions.}
\begin{center}
\resizebox{1.00\hsize}{!}{
\begin{tabular}{l     c       }
\hline
{\sc Galaxy population} & (1) ~~~ $\mathcal{D}$($\tilde{r}$,+$\tilde{s}$) ${\sim}$ $\mathcal{D}$($\tilde{r}$,$-$$\tilde{s}$) \\  
\hline
{\sc Passive}           & 21$^{+12}_{-8}$ \%  \\   
{\sc Star-forming}      & 19$^{+12}_{-8}$ \%  \\
\hline
                        & (2) $\mathcal{D}$($\tilde{r}$,$\tilde{s}$$\ge$0) ${\sim}$ $\mathcal{D}$($\tilde{r}$,$\tilde{s}$$<$0) \\  
\hline
{\sc Passive}           & 4$^{+5}_{-2}$ \%    \\
{\sc Star-forming}      & 15$^{+12}_{-7}$ \%  \\
\hline
                        & (3) ~~~ $\mathcal{D}$($\tilde{r}$,$\tilde{s}$) ${\sim}$ $\mathcal{D}$($\tilde{r}$,$\pm\vert$$\tilde{s}$$\vert$) \\
\hline
{\sc Passive}           & 84$^{+10}_{-13}$ \%  \\ 
{\sc Star-forming}      & 81$^{+11}_{-14}$ \%  \\ 
\hline
\end{tabular}}
\end{center}
{\bf (1) Comparison between the $\mathcal{D}$($\tilde{r}$,$\tilde{s}$)
  and its `reflection' with respect to the kinematic axis
  $\mathcal{D}$($\tilde{r}$,$-$$\tilde{s}$). (2) Comparison between
  the positive $\mathcal{D}$($\tilde{r}$,$\tilde{s}$$\ge$0) and the
  negative side $\mathcal{D}$($\tilde{r}$,$\tilde{s}$$<$0) in the
  kinematic axis of $\mathcal{D}$($\tilde{r}$,$\tilde{s}$). (3)
  Comparison between the $\mathcal{D}$($\tilde{r}$,$\tilde{s}$) and
  its collapsed and symmetrized counterpart
  $\mathcal{D}$($\tilde{r}$,$\pm\vert$$\tilde{s}$$\vert$).}
\label{tab:KS_symm}
\end{table}


As can be shown in table \ref{tab:KS_symm} in all cases the K-S tests
produce significant probabilities that each distribution and its
'reflected' counterpart come from the same parent function, except for
the comparison (2)
i.e. $\mathcal{D}(\tilde{r},\tilde{s}{\ge}0){\sim}\mathcal{D}(\tilde{r},\tilde{s}{<}0)$
of the star-forming population. This comparison produces a low but not
negligible probability of 4$^{+5}_{-2}$\,\% which could be understood
as an indication of some substructure in the population of passive
galaxies.

Following the K-S probabilities shown in table
\ref{tab:KS_highlowlum}, we can conclude that the CPPS distributions
of the two luminosity bins are not statistically distinguishable for
passive and star-forming galaxies. This result is in agreement with
previous results in the literature that only found a distinct
behaviour in the CPPS for the brightest galaxy/galaxies in clusters,
as outline in the following. \citet{Biviano_et_al_1992} found that
galaxies brighter than the magnitude of the third-ranked object,
$m_{3}$, have velocities lower than the average velocity of the
remaining galaxies of the cluster. \citet{ENACS_XI} find that
luminosity segregation in the CPPS is evident only for the ellipticals
that are outside galaxy substructures and which are brighter than
$M_{R}{=}{-}$22. \citet{Goto_2005c} found that only the brightest
$M_{z}{<}{-}$23 cluster galaxies present a significantly smaller
velocity dispersion than the $M_{z}{\ge}{-}$23 galaxies, similar to
what had been found in previous works \citep[e.g.][]{ENACS_IV}.

Given the results of table \ref{tab:KS_highlowlum} which show no
significant differences in the CPPS as a function of luminosity we
plot in figure \ref{fig:all_gal1} all luminosities together. We then
proceed to describe the CPPS distribution of passive galaxies. The
isodensity contours of the CPPS distribution which encloses the bulk
of the passive population presents a `peanut' shape while in the
innermost regions of the CPPS, the isodensity contour presents a
`boxy' shape. The CPPS distribution also shows a tail towards large
radii which gets narrower in velocity space outside the virial region.

With regard to the CPPS distribution of the star-forming population,
the isodensity contours have a convex triangular shape increasing in
separation towards larger radii with a maximum in density around
$\tilde{s}$$\sim$0 and $\tilde{r}$$\gtrsim$1 and they also shows a
small but non negligible population of star-forming galaxies
concentrated at the projected center of clusters $\tilde{r}$$\sim$0
and with low l-o-s velocities
\mbox{$\vert$$\tilde{s}$$\vert\lesssim$0.5}. This population of
star-forming galaxies around ($\tilde{r}$,$\tilde{s}$)$\sim$(0,0)
could be linked with two possible origins: ({\it 1}) it could be a
population of star-forming galaxies which survive to the hostile
environment of cluster virial regions. We discuss this possibility in
subsections \ref{ssec:stripp_intensity} and \ref{ssec:stfm_in_cpd} or;
({\it 2}) these could be galaxy interlopers mixed in projection in the
virial region. In this respect, \citet{Diaferio_et_al_2001} probe, in
a set of simulated clusters, the fraction of galaxies that lie at
physical distances larger than $r_{200}$ as a function of projected
clustercentric radius. At the centers of clusters, the interloper
fraction with respect to the total sample is small, around 10\,\%. The
fraction of red ($B{-}V{>}$0.85) cluster members that are interlopers
is even smaller, $\sim$3\,\%. However, the fraction of blue cluster
galaxies which are galaxy interlopers is higher than 50 per cent.

With respect to the fraction of star-forming galaxies, it presents a
linear trend which goes from 20-30\,\% at the cluster centers
$\tilde{r}$$\approx$0 to a $\sim$70\,\% for $\tilde{r}$$\sim$1. This
central value is in good agreement with a set of works which probes
this quantity in virial regions of massive clusters
\citep{Poggianti_et_al_2006,Popesso_et_al_2007, Finn_et_al_2008}
whereas the fraction at the projected virial radius is even higher
than the fraction of star-forming of $\sim$50\,\% found in the field
regions \citep[e.g.][]{Balogh_et_al_2004,CAIRNS_III}. This higher
fraction comes from the fact that we are showing a luminosity range
which includes galaxies less luminous than these works. In fact,
\citet{Haines_et_al_2007} find that in the lowest density regions the
fraction of star-forming galaxies increases steadily as luminosity
decreases from a fraction of 50\,\% for most luminous galaxies $M_{\it
  r'}{=}{-}$21.5 to a fraction of virtually 100\,\% for the less
luminous bin around $M_{\it r'}{=}{-}$18.

Regarding the radial profiles of velocity dispersion, the star-forming
galaxies show an approximately flat radial profile around $\tilde{s}$=1
up to a radius of $\tilde{r}$$\sim$1 while the passive galaxies
present a mild negative slope in the radial profile which goes from
\mbox{($\tilde{r}$,$\tilde{s}$)$\approx$(0,1)} to
\mbox{($\tilde{r}$,$\tilde{s}$)$\approx$(1,0.7)}. We consider that the
last point at $\tilde{r}$$\sim$1.15 is contaminated by nearby clusters
as mentioned in the beginning of this subsection.

Given that the bootstrap uncertainties in this work are of the order
of 10-15\,\%, we say that the kinematic behaviour of passive and
star-forming galaxies are not distinguishable inside most parts of the
virial region, except for a small decrease in velocity dispersion of
passive galaxies around $R_{P}{=}r_{200}$. The first statement alone
would them suggest that both passive and star-forming galaxies are
approximately virialized inside the virial region, as it was also
pointed out by previous authors \citep[see][and references
  therein]{Biviano&Girardi_2003}, whereas the decrease of velocity
dispersion for passive galaxies around $R_{P}{=}r_{200}$ could be
explained by the contribution of backsplash galaxies. Simulations show
that these galaxies present a more centrally peaked velocity histogram
in contrast with that for the infalling galaxies, which present a
velocity peak well apart from the cluster average l-o-s velocity
\citep[e.g.][]{Gill_et_al_2005}.

It is worth mentioning that \citet{Goto_2005} found that the family
broadly identified as star-forming late-type blue galaxies have a
systematically larger velocity dispersion than that for the family
composed by the passive late-type red galaxies. Specifically, the
largest difference is found between the star-forming population
$\langle{\tilde{s}}\rangle$=1.128 and the passive population
$\langle{\tilde{s}}\rangle$=0.960, when they are splitting them by
their star formation rate (SFR) at SFR(H$\alpha$)=2 M$\odot$
yr$^{-1}$. This difference decreases when the sample is split by
color, where $\langle$$\tilde{s}$$\rangle$=1.085 for the blue
$u'{-}r'{<}$2.22 population and $\langle$$\tilde{s}$$\rangle$=0.961
for the red $u'{-}r'{\ge}$2.22 population. This trend found by
\citet{Goto_2005} suggests that the kinematic behaviour of
star-forming galaxies shows a larger difference with respect to the
passive galaxies when they show a more intense star-formation activity
as also we show in subsection \ref{ssec:stfm_in_cpd}.

\subsection{Clusters from the Millenium Simulation in the CPPS}
\label{ssec:mill_cpps}

Taking into account the distribution of simulated galaxies in the CPPS
depending on the time when they were accreted by their parent cluster,
\citet{Haines_et_al_2012} were able to identify two broad categories
of galaxies, those galaxies accreted at an early epoch and those
accreted after the formation of the core. The first category is well
concentrated at the cluster center and with low l-o-s velocities
suggesting a longer life settled at the cluster core. The second
category is composed by one population of infalling galaxies and
another population of galaxies which have completed their first
pericenter and are now moving outward which are the so-called bound
backsplash galaxies \citep{Mamon_et_al_2004,Gill_et_al_2005,
  Pimbblet_2011}. Therefore, taking advantage of the CPPS to
distinguish the accretion epoch of a specific galaxy depending on its
location in the CPPS and also, considering the clusters are hostile
environments to the star-formation activity, we proceed to study the
similarities and differences between ({\it i}) the CPPS distributions
of passive and star-forming galaxy populations and ({\it ii}) the CPPS
distributions of the accreted galaxy halo population and the
non-accreted galaxy halo populations. We labeled the accreted galaxy
halos as those galaxy halos which have passed inside a certain
clustercentric radius (e.g. $r_{200}$) before an specific cosmic
epoch, for instance a look-back time of one Gyr, and the non-accreted
halo population as those halos which are still outside this radius
before this cosmic epoch. Within this approach, it is assumed that a
star-forming galaxy after it is accreted inside the accretion radius
eventually becomes a passively-evolving galaxy.

For this purpose, cosmic regions centred on 28 massive clusters
(M$_{200}{>}2{\times}10^{14}h^{-1}{\rm M}_{\odot}$) were extracted
from the Millennium simulation \citep{Springel_et_al_2005_Nat}. The 28
clusters were specially selected ({\it 1}) to have clean caustics
without significant contamination from background structures, and
({\it 2}) to have velocity dispersions (for galaxies with
$R_{P}{<}r_{200}$) matching those of our observed cluster sample. In
the simulated sample, there are three clusters which have velocity
dispersions close to that of ABELL\,1185, while the remainder have
velocity dispersions in the range 540-780 km\,s$^{-1}$ as it is shown
in table \ref{tab:SCS} along with the masses and radii of the selected
simulated clusters. We check that the observed and the simulated
samples follow a similar trend in the $r_{200}$-$\sigma_{c}$ and
$M_{200}$-$\sigma_{c}$ relations. The volumes extracted around each
cluster were checked in the l-o-s direction in order to ensure that
for a distant observer viewing along this axis, all galaxies with
line-of-sight velocities within 3000 km\,s$^{-1}$ of the cluster
redshift would be included, enabling us to fully account for
projection effects.

\begin{table}[!ht]
\caption{\bf Properties of the set of simulated clusters.}
\begin{center}
\resizebox{1.00\hsize}{!}{
\begin{tabular}{l              c              c              c    }
\hline
\hline
      ID      &  $\sigma_{c}$($z{=}$0) & $r_{200}$($z{=}$0) &  $M_{200}$($z{=}$0) \\
              & km s$^{-1}$            & Mpc               &   10$^{14}M_{\odot}$  \\ 
(1)           & (2)                   &   (3)             &    (4)              \\     
\hline
    1    &  1034.21    &  2.452    &  17.187   \\  
    2    &  1053.32    &  2.362    &  15.369   \\  
    3    &  1046.45    &  2.084    &  10.560   \\  
    4    &   708.94    &  1.762    &   6.377   \\  
    5    &   755.39    &  1.814    &   6.955   \\  
    6    &   732.22    &  1.964    &   8.831   \\  
    7    &   706.66    &  1.737    &   6.111   \\  
    8    &   769.25    &  1.925    &   8.317   \\  
    9    &   726.60    &  1.965    &   8.844   \\  
   10    &   738.52    &  1.884    &   7.803   \\  
   11    &   656.06    &  1.766    &   6.417   \\  
   12    &   692.45    &  1.815    &   6.976   \\  
   13    &   780.00    &  1.542    &   4.276   \\  
   14    &   634.93    &  1.494    &   3.891   \\  
   15    &   602.94    &  1.510    &   4.011   \\  
   16    &   661.52    &  1.383    &   3.088   \\  
   17    &   689.75    &  1.487    &   3.833   \\  
   18    &   732.29    &  1.529    &   4.166   \\  
   19    &   621.70    &  1.530    &   4.174   \\  
   20    &   659.68    &  1.468    &   3.691   \\  
   21    &   653.14    &  1.463    &   3.654   \\  
   22    &   603.12    &  1.557    &   4.401   \\  
   23    &   559.93    &  1.559    &   4.419   \\  
   24    &   707.28    &  1.528    &   4.163   \\  
   25    &   537.35    &  1.516    &   4.066   \\  
   26    &   714.62    &  1.456    &   3.598   \\  
   27    &   731.48    &  1.469    &   3.698   \\  
   28    &   712.86    &  1.430    &   3.413   \\  
\hline
\end{tabular}}
\end{center}
{\bf (1) ID number, (2) the velocity dispersions are computed
  including all galaxies which would be identified as cluster members
  via a spectroscopic survey within a projected radius
  $R_{P}{=}r_{200}$, (3) the halo radii $r_{200}$ are calculated from
  the $M_{200}$ and (4) $M_{200}$ values are provided by the
  simulation.}
\label{tab:SCS}
\end{table}

\begin{figure}[!ht]
\centering
\resizebox{1.00\hsize}{!}{\includegraphics{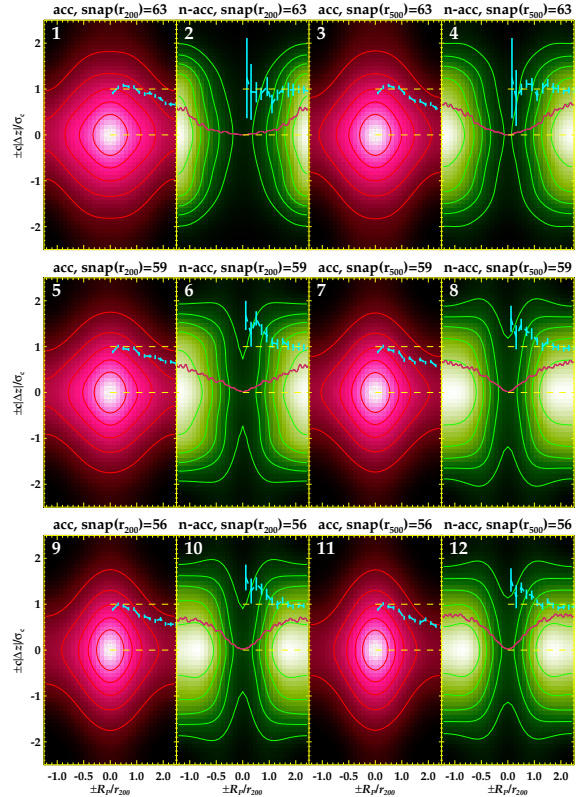}}
\caption{\bf Stacked CPPS of a set of isolated galaxy clusters taken
  from the Millenium Simulation \citep{Springel_et_al_2005_Nat}. The
  simulated population is split in two categories depending on if the
  simulated galaxy was accreted (odd panels for accreted one) or not
  (even panels for non-accreted one) and for two different radii where
  it is assumed the galaxy is accreted $r_{acc}{=}r_{200}$ (two
  columns at the left) and $r_{acc}{=}r_{500}$ (two columns at the
  right). See definition of accreted and non-accreted galaxy in the
  first paragraph of section \ref{ssec:mill_cpps}. From top to bottom,
  the corresponding accretion epochs are the current cosmic epoch
  (snap($r_{acc}$)=63), a look-back time of $t_{LB}$=1.13 Gyr
  (snap($r_{acc}$)=59) and $t_{LB}$=2.09 Gyr (snap($r_{acc}$)=56). In
  these graphs, the solid line shows the radial fraction of
  non-accreted galaxy halos. The rest of elements of the figure are
  defined in the same way as figure \ref{fig:all_gal1}.}
\label{fig:CPD_Millenium}
\end{figure}

The Millennium simulations \citep{Springel_et_al_2005_Nat} cover a
($500~h^{-1}{\rm Mpc}$)$^{3}$ volume, and provide catalogs of dark
matter halos and galaxies, the latter based on the semi-analytic
models ({\sc galform}) of \citet{Bower_et_al_2006}, in which peculiar
velocities, positions, absolute magnitudes and halo masses are given
at 63 snapshots covering a look-back time range from $t_{LB}$=13.57
Gyr to nowadays. The galaxy l-o-s velocities and sky positions for the
28 systems were stacked, setting the origin at the average l-o-s
velocity c$z_{c}$ and at the cluster center,
respectively. Normalization by the velocity dispersion $\sigma_{c}$
and the virial radius $r_{200}$, respectively, was then
performed. Figure \ref{fig:CPD_Millenium} shows the stacked CPPSs for
simulated galaxies with $M_{K}{<}{-}$22.60 \citep[which corresponds to
  $M^{\ast}_{K}{+}$2.36,][]{Merluzzi_et_al_2010} in the 28 simulated
clusters, split into two categories depending on the accretion epoch
and for two different values of accretion radius, $r_{acc}{=}r_{200}$
(left) and $r_{acc}{=}r_{500}$ (right).

Assuming we do not find luminosity segregation in the CPPS for the
whole cluster galaxy sample, it is fair to compare the observed sample
with this sample of simulated galaxies showing a $K$-band luminosity
range down to $M^{\ast}_{K}{+}$2.36. This simulated sample is
comparatively inside the observed $r'$-band luminosity range which
reaches down to $M^{\ast}_{\it\,r'}{+}$3.21. We show the three
snapshots corresponding to the current cosmic epoch and the values of
look-back time closer to 1 Gyr and 2 Gyr; these are snap=63
(nowadays), snap=59 ($t_{LB}$=1.13 Gyr) and snap=56 ($t_{LB}$=2.09
Gyr). We choose this time step of \mbox{$\sim$1 Gyr} because it is a
typical value for the crossing time for a galaxy in a massive cluster
\citep{Boselli&Gavazzi_2006}. Furthermore, this time-scale is of the
same order of magnitude than the time-scale of galaxy evolution of
optical colors \citep{Kennicutt_1998}, such as $u'{-}r'$ which is of
interest in the study of the CPPS distribution of star-forming
galaxies depending on their intensity of star formation, as shown in
subsection \ref{ssec:stfm_in_cpd}.


In a first inspection of figure \ref{fig:CPD_Millenium}, one can
notice that there is a qualitative agreement between the CPPS
distribution of the observed galaxy populations in figure
\ref{fig:all_gal1} and the simulated halo populations in figure
\ref{fig:CPD_Millenium}. This suggests that the attempt to link the
passive population with the accreted population and the star-forming
population with the non-accreted population is a good approach.

With respect to the CPPS distribution of the accreted population (odd
panels in figure \ref{fig:CPD_Millenium}), the isodensity contours
present a round/elliptical shape in the virial region. In addition,
the population inside the projected virial radius $R_{P}{<}r_{200}$ is
comparatively larger than the population outside the projected virial
radius $R_{P}{>}r_{200}$ as ({\it a}) one goes back in time
i.e. longer look-back time of accretion and ({\it b}) larger accretion
radius are considered i.e. $r_{acc}{=}r_{200}$ instead of
$r_{acc}{=}r_{500}$. In fact, for an accretion radius of $r_{200}$ and
when the accretion epoch is set to the current cosmic epoch, we have
the largest population outside the projected virial radius
$R_{P}{>}r_{200}$ as compared to the population inside the projected
virial radius $R_{P}{<}r_{200}$. Regarding the radial profile of
velocity dispersion of accreted population, it shows a monotonic
decrease towards large radii in all shown snapshots, which is in
agreement with the results that find a more centrally peaked velocity
distribution for the backsplash population
\citep{Gill_et_al_2005}. This profile also shows a small dip at
$\tilde{r}{\sim}$0.

Regarding the CPPS distribution of non-accreted population (even
panels in figure \ref{fig:CPD_Millenium}), it presents a maximum
around ($\tilde{r}$,$\tilde{s}$)$\sim$(1,0) which is relatively round
for a current cosmic epoch of accretion and an accretion radius of
$r_{acc}{=}r_{200}$ while it becomes more `boxy' for longer look-back
times of accretion and smaller accretion radii. The non-accreted
population occupies more inner parts of virial region when it is
considered a smaller accretion radius ($r_{500}$) and longer look-back
times of accretion. Regarding the radial profile of velocity
dispersion (dashed line with uncertainty bars), it changes from a
flatter profile when the accretion epoch is nowadays to a radial
profile showing an increase around $\tilde{r}{\sim}$0 for earlier
accretion epochs. The radial fraction of the non-accreted population
(solid line in figure \ref{fig:CPD_Millenium}) goes from a very low
value lower than 3\,\% at the projected cluster center to around
60\,\% at $R_{P}{\sim}$1.25$r_{200}$ in the case the accretion radius
$r_{acc}{=}r_{200}$ and it reaches around 70\,\% in the case the
accretion radius $r_{acc}{=}r_{500}$. We point out the presence of
non-accreted haloes inside the virial radius $R_{P}{<}r_{200}$ at all
shown snapshots and in some cases, even in the low velocity interval
$\tilde{s}$$\sim$0.

\section{Discussion: Similarities and differences between the CPPS distributions of the galaxy populations and the halo populations}
\label{sec:discussion}

In the attempt to link the distribution of accreted halos with the
passive galaxies and the distribution of non-accreted halos with the
star-forming galaxies, we consider the distribution of
accreted/non-accreted galaxy halos in the case of snap($r_{200}$)=63
as the most similar distribution to the passive/star-forming
galaxies. We are taking into account that this is the snapshot where
({\it i}) the contribution of the accreted population outside the
projected virial radius $R_{P}{>}r_{200}$ to the whole accreted
population is the largest of the cases shown here and the most similar
to the contribution of the passive population outside the projected
virial radius $R_{P}{>}r_{200}$ to the whole passive population and
({\it ii}) the radial profile of velocity dispersion and the range of
l-o-s velocities of non-accreted halos inside the virial radius
$R_{P}{<}r_{200}$ are the mos similar to those ones of star-forming
galaxies. On the other hand, where we do not find a good match is in
the fraction of non-accreted population near the projected center
which is lower than 3\,\% and is clearly lower than the corresponding
fraction of star-forming galaxies of 20-30\,\%. This difference could
be explained if the star-forming galaxies inside the physical virial
region $r{<}r_{200}$ retain a remaining star formation activity. This
possibility is discussed in subsections \ref{ssec:stripp_intensity}
and \ref{ssec:stfm_in_cpd}. Further, the non-accreted population
covers a larger l-o-s velocity range than the star-forming
galaxies. Gas stripping can be invoked to alleviate this
difference. This will be discussed in subsection
\ref{ssec:stripp_intensity}. Although there is clearly a qualitative
similarity between the CPPS distributions of the accreted population
and the passive population especially in the case of
snap($r_{200}$)=63, they are quantitatively
distinguishable. Specifically, a K-S test applied to both CPPS
distributions i.e the simulated case of snap($r_{200}$)=63 and the
observed passive population produces a negligible probability that
these ones come from the same parent function i.e. $\mathcal{D}{\rm
  (accreted)}{\sim}\mathcal{D}{\rm (passive)}\lesssim0.18\%$.

At this point, we outline two caveats in the one-to-one identification
of the accreted population with the passive population and the
non-accreted population with the star-forming population. First, the
non-accreted galaxy population is composed by both star-forming and
passive galaxies in a mix determined by the typical galaxy population
from the field. In the rarefied field, \citet[][]{Haines_et_al_2007}
found the fraction of passively-evolving galaxies is a strong function
of luminosity, decreasing from 50 per cent for
$M_{\it\,r'}{\lesssim}{-}$21 to zero by
$M_{\it\,r'}{\sim}{-}$18. Secondly, the accreted galaxy population, as
we see later in subsection \ref{ssec:stripp_intensity}, does not have
to be necessarily solely identified with the passive galaxy
population. In any case, it is observationally found that even the
regions with the highest volume densities of galaxies show a fraction
of $\sim$30 per cent of star-forming galaxies regardless of luminosity
\citep[][see fig. 3]{Haines_et_al_2007}.

\begin{figure}[!ht]                                         
\centering 
\resizebox{1.00\hsize}{!}{\includegraphics{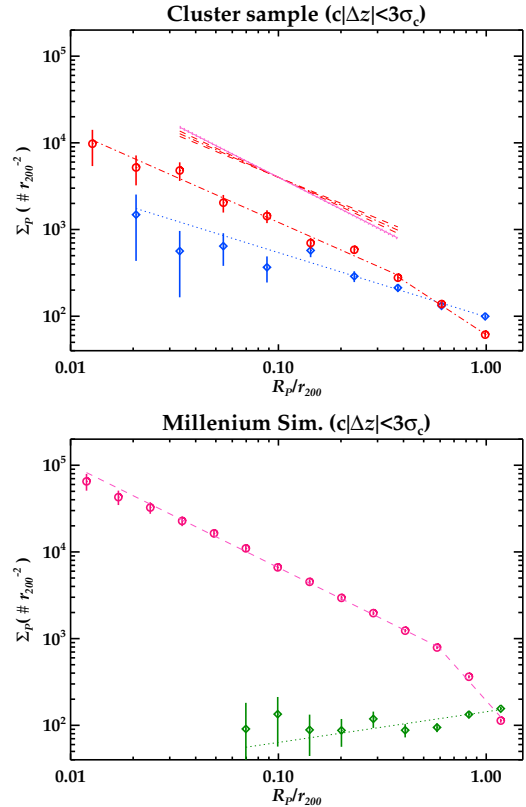}}
\caption{\bf Top graph: Radial profile of the projected density (in
  natural units) of cluster galaxies. Circles correspond to passive
  galaxies while diamonds correspond to star-forming galaxies,
  including poissonian uncertainties as vertical bars. Dot-dashed and
  dotted lines through the data are power-law fits to the passive and
  the star-forming population, respectively. Solid and dashed lines
  above the data, including uncertainties in slopes shown as tilted
  lines around them, are respectively the fits to the simulated
  accreted and observed passive populations. Bottom graph: Radial
  profile of the projected density of accreted (circles) and
  non-accreted (diamonds) galaxy halos, including poissonian
  uncertainties. Dashed and dotted lines are power-law fits to the
  simulated accreted and non-accreted populations, respectively.}
\label{fig:radprof}
\end{figure}

\begin{figure}[!ht]
\centering
\resizebox{1.00\hsize}{!}{\includegraphics{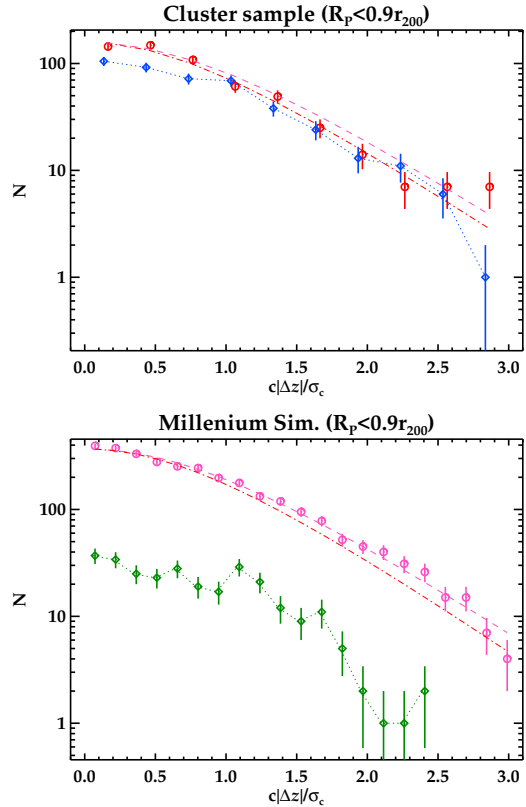}}
\caption{\bf Top panel: Histograms of l-o-s velocity (in natural
  units) of passive (circles) and (diamonds) star-forming
  galaxies. Bottom panels: Histograms of l-o-s velocity (in natural
  units) of accreted (circles) and (diamonds) non-accreted galaxy
  halos. The dashed-dot and dashed lines are fits (see text for
  details) to the velocity distributions of passive and accreted
  population, respectively. In both panels, the fitted functions are
  scaled to the central value $N_{0}$ to make easier the comparison
  between them. Dotted lines are just joining the data points of
  star-forming and non-accreted velocity distributions.}
\label{fig:velhist}
\end{figure}

On the comparison of the accreted population and the passive one, the
most striking difference is the shape of the isodensity contours where
they are elliptical or well rounded in the case of the accreted
population and clearly `boxy' or with a characteristic `peanut"
shape for the observed distribution. Assuming the distribution in the
CPPS is a combination between the radial profile of projected density
and the velocity distribution of galaxies, we proceed to derive these
two observables in order to know where the differences come from. For
both data and simulations, the radial profiles of projected density
are shown in figure \ref{fig:radprof} while the velocity histograms
are shown in figure \ref{fig:velhist}.

                                                     
By comparing the radial profiles of projected density for the observed
and simulated populations shown in figure \ref{fig:radprof}, we find
that both populations are well characterized by a broken power law
\mbox{$\Sigma_{P}~{\propto}$~$\tilde{r}$$^{~\alpha}$} within the
uncertainties. The star-forming population seems to follow a simple
power-law behaviour along the whole range of radii under consideration
i.e. the virial region, while the passive population shows a break at
$R_{P}{\sim}$0.4$r_{200}$. In the case of the simulated clusters, the
projected density profiles of the non-accreted population present
approximately a simple power-law behaviour along the virial region and
the accreted population shows a break at
$R_{P}{\sim}$0.5$r_{200}$. Specifically, the fits to radial profiles
produce the following logarithmic slopes which are shown in table
\ref{tab:slope_clust}.


\begin{table}[!ht]
\caption{\bf Logarithmic slopes of the radial profiles of projected
  density for the galaxy and halo populations studied here.}
\begin{center}
\resizebox{1.00\hsize}{!}{
\begin{tabular}{lll}
\hline
                  & Cluster sample                   & \\
\hline                                               
                  & {\sc Passive}                    &  {\sc Star-forming}                 \\                    
$\tilde{r}$$<$0.4 & $\alpha$ = $-$1.060 $\pm$ 0.065  &   $\alpha$ = $-$0.742  $\pm$  0.055 \\   
$\tilde{r}$$>$0.4 & $\alpha$ = $-$1.57  $\pm$ 0.12   &                                     \\  
\hline                                               
                  & Millenium simulation             &   \\
\hline                                               
                  & {\sc Accreted}                   &   {\sc Non-accreted}                \\  
$\tilde{r}{<}$0.5 & $\alpha$ = $-$1.187 $\pm$ 0.022  &   $\alpha$ = $$0.354  $\pm$  0.081  \\   
$\tilde{r}$$>$0.5 & $\alpha$ = $-$2.695 $\pm$ 0.094  &   \\   
\hline
\end{tabular}}  
\end{center}
\label{tab:slope_clust}
\end{table}


As one can see in table \ref{tab:slope_clust}, the radial profile of
the passive population shows an inner slope very close to
$\alpha{=}-$1, which implies a flat radial profile in number density
(i.e. the number of galaxies per unit radius remains constant) and
this seems to be the main cause of the `boxy' and/or `peanut' shape of
isodensity contours of the passive population. Inspecting the top
graph of figure \ref{fig:radprof}, the inset lines for the comparison
of slopes show that the radial profile slope of the accreted
population is marginally steeper than that for passive galaxies
assuming the size of slope uncertainties. Further, the bias produced
by uncertainty in cluster centering is able to reduce the logarithmic
slope of radial profile of accreted halos enough to match the slope of
the observed radial profile of passive galaxies in the innermost part
of the virial region. We discuss this issue in depth in subsection
\ref{ssec:appendixC}.

As referred to the density profile of non-accreted haloes, the
relatively low density of non-accreted haloes with respect to the
star-forming population inside the virial radius suggests that we need
a contribution of accreted haloes surviving inside the virial radius
with a remaining of star formation as we discuss in subsections
\ref{ssec:stripp_intensity} and \ref{ssec:stfm_in_cpd}.

With regard to the distributions of l-o-s velocities, we fit a
function in figure \ref{fig:velhist} of the form of an exponential
with the negative branch of a vertical hyperbola as argument i.e.:

$N_{0}${\it exp}$\biggl\{s_{1}\biggl(~1-\biggl( 1+(\tilde{s}/s_{0})^{2}\biggr)^{1/2}~\biggr)\biggr\}$ 

obtaining for the passive population: \\
$N_{0}{=}$159$\pm$11, $s_{1}{=}$2.4$\pm$1.7 and $s_{0}{=}$1.16$\pm$0.54.

For the accreted population we obtain: \\ 
$N_{0}{=}$369$\pm$11, $s_{1}{=}$3.00$\pm$0.96 and $s_{0}{=}$1.43$\pm$0.29.


Taking into account the uncertainties in velocity histograms, it is
not easy to establish a connection between the l-o-s velocity
distributions of passive galaxies and the accreted halos but it is
significant that both velocity distributions approximately follow an
exponential law $N{\approx}exp$(${-}\frac{s_{1}}{s_{0}}~\tilde{s}$)
for an intermediate velocity range
0.5${\lesssim}\tilde{s}{\lesssim}$2.5. Although, the parameters
obtained from fitting to the velocity distributions of the passive and
accreted populations are noticeably different, the shapes of fitted
functions are compatible within uncertainties, as can be seen in
figure \ref{fig:velhist}. It is clear that these two parameters
present some correlation between them.

On the relation of the shape of the velocity distribution with the
type of galaxy orbits, \citet{Merritt_1987} derived the velocity
distribution of galaxies in three extreme cases where the orbits are
circular, radial or isotropic; finding a more cuspy distribution in
the radial case, a bell-shaped distribution in the isotropic case and
flat-topped distribution for the circular \citep[see
  also][]{Van_der_Marel_et_al_2000}. Although, a number of studies
found a very low or negligible anisotropy at the center of the
clusters \citep[e.g.][]{ENACS_XIII,Hwang&Lee_2008,Wojtak&Lokas_2010},
it is worth noting that the contribution of galaxy substructures is
non negligible even at the very center \citep[][fig. 3]{ENACS_XI},
which complicates the relation between the velocity distribution and
the kinematic state of galaxy population. In addition, even in the
case of large samples of clusters, the velocity histograms of clusters
show differences from Gaussian curves
\citep{Blackburne&Kochanek_2012}.

From the comparison of the radial profiles of the projected density
and the velocity distributions between the passive and the accreted
populations, we can conclude that the differences in the isocontour
shape between the passive population and the accreted one mainly come
from the difference on the inner slope of the radial profile of
projected density.

\subsection{Effects of uncertainties from cluster centering in the
  radial profile of projected density and the uncertainties of cluster
  average velocity in the velocity distribution.}
\label{ssec:appendixC}

The uncertainties in the determination of cluster center produce an
unavoidable bias in the estimation of the central value of the profile
of projected density toward lower values, such as it was pointed by
\citet{Beers&Tonry_1986}. On the comparison of the radial profiles of
projected density between data and simulations, we have to take into
account this bias. In order to estimate the importance of this bias
depending on the uncertainty size, we take the set of simulated
clusters as fiducial models and include the uncertainties of cluster
centering and also, on the average l-o-s velocity of the cluster as it
is described just below.

For each cluster labeled as $j$, we add uncertainties
$\delta\tilde{r_{j}}$ to the actual clustercentric radii
$\tilde{r}_{i}$ of galaxy labeled as $i$ to obtain the `observed'
radius $\tilde{r}_{i}^{\ast}$ using the following expression derived
from simple trigonometric relations:

$(\tilde{r}_{i}^{\ast})^{2} = (\tilde{r}_{i})^{2} + (\delta\tilde{r_{j}})^{2}
-2 \cdot \tilde{r}_{i} \cdot \delta\tilde{r}_{j} \cdot cos(\theta_{j}-\theta_{i})$ 

with $\delta\theta_{ji}\equiv\theta_{j}-\theta_{i}$ being the angle
difference between $\theta_{j}$, the polar angle of vector
$\delta\tilde{r}_{j}$, and $\theta_{i}$, the polar angle of vector
$\tilde{r}_{i}$. Assuming that $\theta_{i}$ is statistically
independent of $\tilde{r}_{i}$, we can randomly generate
$\delta\theta_{ji}$ following an uniform distribution $\mathcal{U}$ in
all directions,
i.e. $\delta\theta_{ji}\thicksim\mathcal{U}_{j}~[~0,2\pi~]$ and the
uncertainty $\delta\tilde{r}_{j}$ can be generated following the
modulus of a normal distribution
i.e. $\delta\tilde{r}_{j}\thicksim\parallel\mathcal{N}(0;\sigma_{\tilde{r}}[~r_{200}~])\parallel$. In
addition, we also test the effect of the uncertainty in the estimation
of the average l-o-s velocity of the cluster $\delta\tilde{s}_{j}$
over the actual cluster-frame l-o-s velocity $\tilde{s}_{i}$ of galaxy
$i$ as:

$\tilde{s}_{i}^{\ast} = \tilde{s}_{i} + \delta \tilde{s}_{j}$

\begin{figure}[!ht]
\centering
\resizebox{1.00\hsize}{!}{\includegraphics{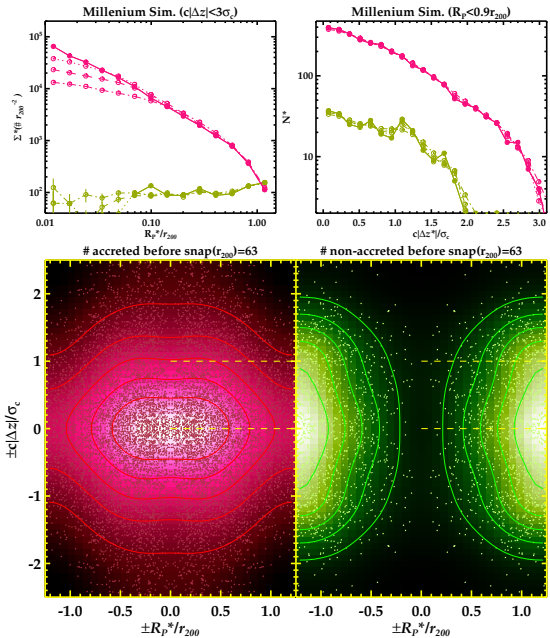}}
\caption{\bf Illustration of the effect of adding uncertainties to the
  cluster centering (top left and bottom panels) and to the average
  l-o-s velocity (top right panel) in the case of snap($r_{200}$)=63
  (top left panel in figure \ref{fig:CPD_Millenium}). Top left panel:
  Radial profiles of projected density when it is added 0.05$r_{200}$
  (dotted line), 0.1$r_{200}$ (dashed line) and 0.2$r_{200}$
  (dot-dashed line) of uncertainty in cluster centering from top to
  bottom, respectively. Top right panel: Velocity distributions when
  it is added 0.05$\sigma_{c}$ (dotted line), 0.1$\sigma_{c}$ (dashed
  line) and 0.2$\sigma_{c}$ (dot-dashed line) of uncertainty in
  cluster average l-o-s velocity. In both top graphs, the filled
  circles joined with solid lines are the `unperturbed'
  profiles. Bottom graphs: Stacked CPPS of the set of simulated galaxy
  clusters in the case of snap($r_{200}$)=63 including a 0.2$r_{200}$
  of uncertainty just in cluster centering. The rest of elements of
  the figure are defined in the same way as figure
  \ref{fig:CPD_Millenium}.}
\label{fig:radprof_velhist_Millu}
\end{figure}

We model this effect assuming a normal distribution for the
uncertainty $\delta\tilde{s}_{j}$ to compute the `observed'
cluster-frame l-o-s velocity $\tilde{s}_{i}^{\ast}$ i.e. $\delta
\tilde{s}_{j}\thicksim\mathcal{N}(0;\sigma_{\tilde{s}}~[~\sigma_{c}~])$. The
result of 100 trials, using these inputs for both the radial profile
of projected density and the velocity histogram, is summarized in
figure \ref{fig:radprof_velhist_Millu}.


As one can see in figure \ref{fig:radprof_velhist_Millu} for the case
of the radial profile of the projected density, the fact of not
exactly centering the profile produces a systematic underestimation of
the central density. Regardless the importance of the bias, this is
induced in a differential way along the radial range i.e. the bias is
more intense in the innermost regions with a radius scale of the order
of the average uncertainty. This produces a more pronounced curvature
in the radial profile. Comparing the `observed' radial profile of
accreted halos including centering uncertainties with the radial
profile of passive galaxies, we can derive that a relative uncertainty
of around 0.2$r_{200}$ is enough to get the inner slope of radial
profile of passive galaxies taking as parent distribution the radial
profile of accreted halos. With respect to the l-o-s velocity
histogram; it is required large values of uncertainty in the average
l-o-s cluster velocity, well larger than a 0.2$\sigma_{c}$, to produce
a noticeable departure from the parent distribution. One of the
simplest conclusions that we can derive from this computational test
is that the induced bias is only significant in the case when the
uncertainty is of the order of the typical scale of variation of the
function under study, as it was pointed out by \citet{ENACS_VII}.

The uncertainties involved in the center and the average l-o-s
velocity of clusters derived through the caustic method present an
standard deviation of around 100\,km\,s$^{-1}$ in l-o-s velocity and
around 100${h^{-1}}$\,kpc in the projected center on the sky (Diaferio
2013, private communication) but the departure from actual center and
average l-o-s velocity can reach values of 0.5${h^{-1}}$\,Mpc on the
sky and 400\,km\,s$^{-1}$ along the line of sight, respectively
\citep{Serra&Diaferio_2013}. Moreover, the comparison of caustic
centers with the X-ray emission peaks of the intracluster medium (ICM)
produces differences of the order of 150\,kpc \citep{CIRS}. In one
hand, comparing these centering uncertainties with the values of
virial radius in our cluster sample $r_{200}{\sim}$1.2-2.3 Mpc (see
table \ref{tab:CS}), we can conclude that these values of uncertainty
could explain the difference between the inner logarithmic slope of
passive galaxies and the corresponding one of the accreted halos. This
allows to assume that the bulk of passive galaxies in the innermost
parts of virial region are accreted galaxy halos. \citet{ENACS_VII}
simulate the effect of centering uncertainties using a NFW profile
\citep{NFW} with an inner slope of $\alpha{\sim}-$0.66 and do not find
a significant decrease in the central projected density. Also, they
find that the NFW profiles fitted to individual clusters, according to
the likelihood ratio statistic, are generally not statistically
distinguishable from the prototypical core profile, the King
profile. Further, a King profile is favoured for the observed radial
profile of the composite cluster sample at the innermost radii. As
\citet{ENACS_VII} argue the centering uncertainties tend to destroy
the cusp in composite profiles. We add that this effect is stronger in
cuspy profiles with a pronounced logarithmic slopes as indicate the
results shown here. This differential behaviour between core profiles
and cuspy profiles becomes more evident when the centering
uncertainties are of the order of the core radius. In short, the
differences in the effect of centering uncertainties between the test
made by \citet{ENACS_VII} and our test come from the fact that we test
with a clearly more cuspy profile ($\alpha{\sim}-$1.19) than their
test which was performed with a core profile. At this point, we have
to mention that there are noticeable differences in the inner slope
between different semi-analytic models \citep{Budzynski_et_al_2012}
and the effect of cluster centering uncertainty is greater for higher
values of the inner slope. On the other hand, comparing the typical
uncertainty of the average l-o-s cluster velocity c$z_{c}$ with the
range of cluster velocity dispersions
$\sigma_{c}{\sim}$500-1000\,km\,s$^{-1}$, this produces around a
0.1-0.2$r_{200}$ relative uncertainty, which is not enough to clearly
change the cluster velocity distribution, as one can be seen in figure
\ref{fig:radprof_velhist_Millu}.

What seems more significant, the appearance of the CPPS distribution
of simulated galaxy haloes in the case of snap($r_{200}$)=63 (see
bottom panel in fig. \ref{fig:radprof_velhist_Millu}) is quite similar
to the CPPS isodensity contours of the passive population, when it is
added an uncertainty of 0.2$r_{200}$ just in cluster centering.
Comparing figures \ref{fig:all_gal1} and
\ref{fig:radprof_velhist_Millu}, it is easy to see that the inner
isocontours of the accreted population become similar to those ones of
passive galaxies showing this characteristic `peanut' shape and
specifically, the isocontour of highest density presents a relatively
``boxy' shape. The main difference between isodensity contours of the
accreted population with centering uncertainty and the passive
population is observed at $R_{P}{>}$0.7$r_{200}$, where the accreted
population has two `open' isocontours while the passive population has
three `open' isocontours. This difference likely comes from the
  fact that there is an important contribution of passive galaxies in
  the non-accreted population as it is observed in the rarefied field
  \citep{Haines_et_al_2007}. In addition, the simulated clusters were
  specially selected to be isolated clusters with a very low or
  negligible contamination from nearby clusters, whereas the set of
  real clusters shows contribution from neighboring clusters in their
  surroundings (see figure 1 in Paper I).  This two facts will
  comparatively increase the CPPS density of passive galaxies around
  $R_{P}{\sim}r_{200}$ and consequently this will modify the shapes of
  contours of lowest density for accreted halos changing them to
  `open' contours.

\subsection{Gas stripping in the modulation of the CPPS distribution of galaxy populations}
\label{ssec:stripp_intensity}

Motivated by the hint provided by the scarcity of star-forming
galaxies near the projected center $\tilde{r}$$\sim$0 for relatively
high l-o-s velocities $\tilde{s}$$\gtrsim$0.5, we proceed to explore
the influence of the ram-pressure stripping in shaping the
distribution of galaxies in the CPPS. We compute the `intensity' of
this environmental effect throughout the CPPS as the ratio of the
ram-pressure $P_{\rm ram}$ over the anchoring gas pressure of a galaxy
$\Pi_{gal}$ \citep{Gunn&Gott_1972}. We model the gas density profile
with the classical $\beta$-profile
\citep{Cavaliere&Fusco-Femiano_1976}:

\begin{displaymath}
\left. \begin{array}{l} 
\rho_{\rm ICM}(r) = \rho_{0} \left[ 1 + \left( \frac{r}{R_c} \right)^{2} \right]^{-\frac{3}{2}\beta} \sim \\
\sim \rho_{0} \left[ 1 + \left( {\rm c} \tilde{r} \right)^{2} \right]^{-\frac{3}{2}\beta} = \rho_{0} ~{\rm \tilde{K}}(\tilde{r};~c,\beta) = \rho_{0} ~{\rm \tilde{K}}(\tilde{r})
\end{array} \right.
\end{displaymath}

with $\rho_{0}$ being the central ICM density, $R_{\rm c}$ the core
radius of the $\beta$-profile and
c$\equiv$($r_{200}$/$R_{\rm\,c}$)($\pi$/2), where we are assuming that
the three-dimensional radius $r$ scales with scatter as
\mbox{$r{\sim}\frac{\pi}{2}R_{P}$}, in order to account for projection
effects.

\begin{displaymath}
\left. \begin{array}{r} 
P_{\rm ram} = \rho_{\rm ICM} \times v_{\rm gal}^{2} \sim \rho_{0} {\rm \tilde{K}}(\tilde{r}) \times \tilde{s}^{2} (3 \sigma_{c}^{2}) \\
\Pi_{\rm gal} \sim \Pi_{\rm MW}
\end{array} \right \} \Longrightarrow
\end{displaymath}

\begin{equation} \label{eq:strippeff}
~~\Longrightarrow~~ 
\eta = \frac{P_{\rm ram}(\tilde{r},\tilde{s})}{\Pi_{\rm MW}} ~; ~~ \tilde{s}^{2} ~ {\rm \tilde{K}}(\tilde{r}) = 
\eta~ \left( \frac{\Pi_{\rm MW}}{3 \rho_{0} \sigma_{c}^{2}} \right) 
\end{equation}

We take $v_{\rm gal}^{2}{\sim}$~$\tilde{s}$~(3$\sigma_{c}^{2}$) as a
proxy of galaxy velocity in the cluster frame and $\Pi_{\rm MW}$ as
the anchoring gas pressure in the Milky Way as a representative value
for galaxies. Finally, as a fiducial example, we take the derived
values for cluster CL+0024 \citep{Treu_et_al_2003}: $R_{c}{=}$60\,kpc,
$r_{200}{=}$1.7\,Mpc, $\beta{=}$0.475,
$\rho_{0}{=}$3$\cdot$10$^{14}$\,M$_{\odot}$\,Mpc$^{-3}$,
\mbox{$\sigma_{c}$=911\,km\,s$^{-1}$} and \mbox{$\Pi_{\rm
    MW}$=2.1$\cdot$10$^{-12}$ N m$^{-2}$} as a reference for the
restoring gravitational force per unit area (units of pressure) for a
galaxy like the Milky Way (see also
\citealt[][fig. 2]{Abadi_et_al_1999} and
\citealt[][fig. 18]{Boselli&Gavazzi_2006}). For comparison,
\citet{Treu_et_al_2003} propose a `stripping radius' of 0.5-1 Mpc for
this galaxy cluster CL+0024, but this approach is not taking into
account the kinematic axis. Further, \citet{Boselli&Gavazzi_2006}
explore the dependence of the HI deficiency parameter of Virgo A
galaxies on the synthetic observable
(1/$\theta$)\,$\times$\,v$_{{l}\cdot{o}\cdot{s}}^{2}$, which provides
a broad one-dimensional trend of ram-pressure intensity, with $\theta$
the projected cluster radius in angular units and
$v_{{l}\cdot{o}\cdot{s}}$ the cluster-frame l-o-s velocity.

\begin{figure}[!ht]
\centering
\resizebox{1.00\hsize}{!}{\includegraphics{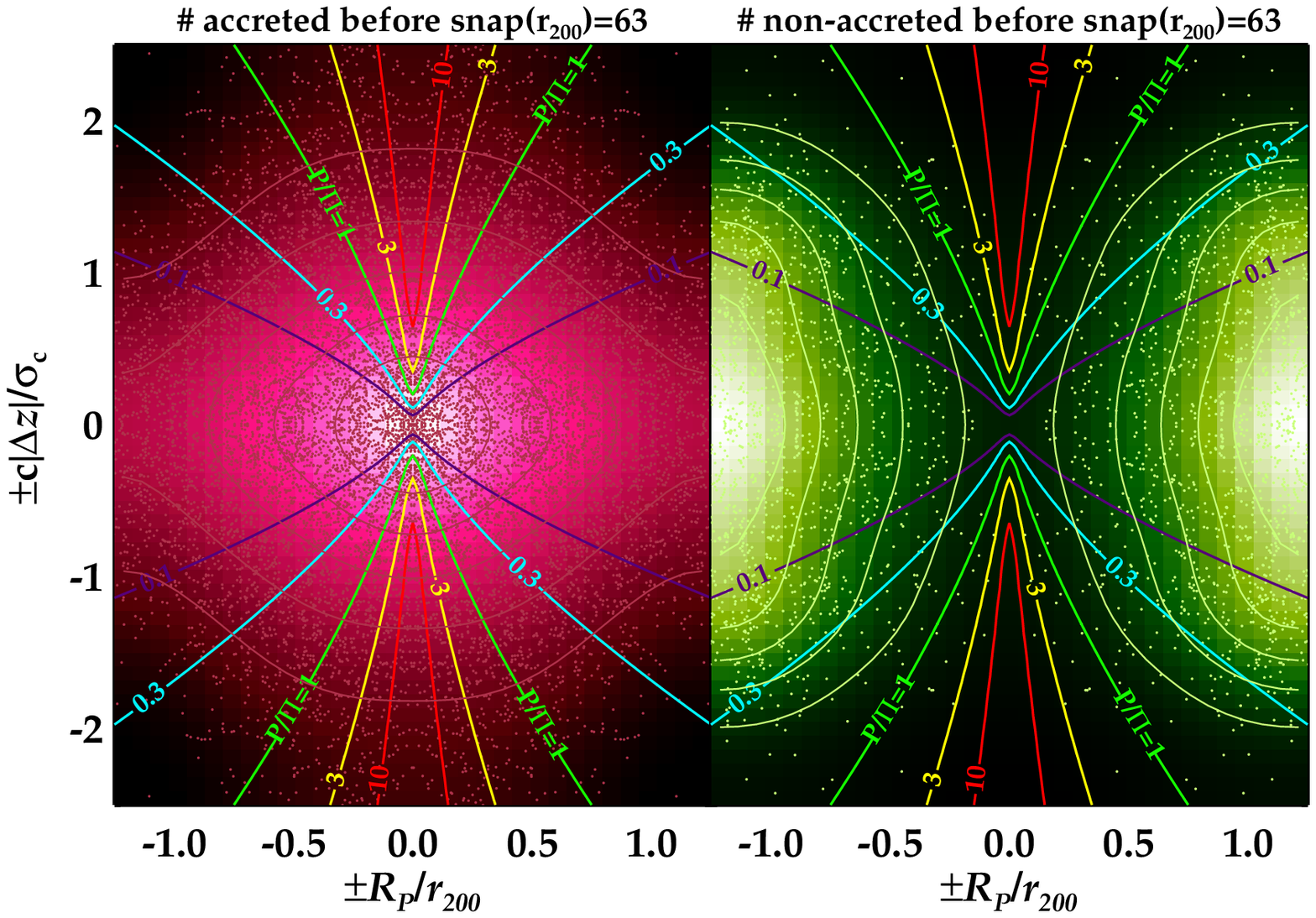}} \\ 
\resizebox{1.00\hsize}{!}{\includegraphics{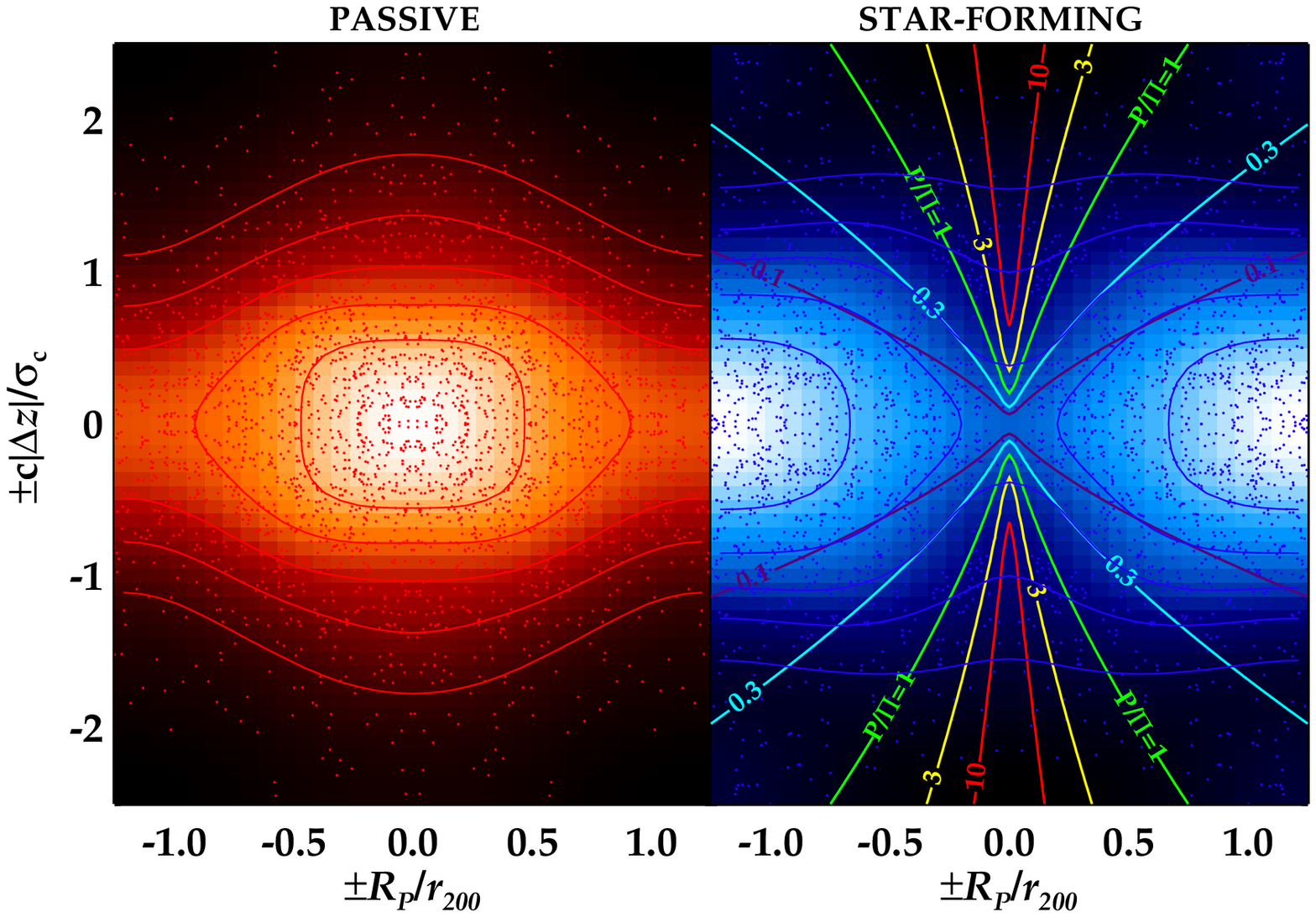}}
\caption{\bf Stripping intensity. Isocontours of stripping intensity
  $\eta{=}P_{\rm ram}{/}\Pi_{\rm gal}$ overplotted in the CPPSs of the
  halo populations in the case of snap($r_{200}$)=63 (top) and the
  galaxy populations (bottom). Each contour is labeled by the ratio of
  ram pressure over the anchoring pressure of gas in the Milky Way
  $\eta$ (see equation \ref{eq:strippeff} and corresponding text for
  details) with this set of values \mbox{$\eta$ = [ 0.1, 0.3, 1, 3, 10
  ]}. The rest of elements of the figure are defined in the same way
  as figure \ref{fig:high-lum_gal1}.}
\label{fig:stripp_eff_fig}
\end{figure}

We make note of a number of caveats. First, we have to keep in mind
that we are probing the effects of an environmental phenomenon which
depends on at least six environmental variables (three spatial and
three velocity components) in a projected two-dimensional
diagram. Also, despite the fact that we apply a scaling factor of
\mbox{$v_{gal}{{\sim}}\sqrt{3}\sigma_{c}$} to account for the galaxy
velocity, a certain amount of scatter seems unavoidable. In the same
way, the ICM density is assumed to be dependant on the
three-dimensional radius $r$ whereas we only have the projected
component $R_{P}$, so we apply an average scaling of
\mbox{$r{\sim}\frac{\pi}{2}R_{P}$}. Furthermore, we are stacking in
the CPPS galaxy distributions, clusters with a range of velocity
dispersions $\sigma_{c}{\sim}$500-1000\,km\,s$^{-1}$, and a collection
of galaxies with different anchoring pressures \citep[][their
  fig. 18]{Boselli&Gavazzi_2006}. Because of these handicaps, our
motivation is to analyse the influence of ram-pressure stripping in
modulating the CPPS galaxy distribution in a qualitative way, more
than to compute a precise shape of the $\eta$ function. Nevertheless,
the comparison produces conspicuous results, as one can see in figure
\ref{fig:stripp_eff_fig}, where we overplotted iso-contours with a
representative set of values of stripping intensity in both the CPPS
of real clusters and simulated clusters. This set of values attempts
to account for ({\it i}) the range of variation of $\eta$ due to the
variation of physical variables involved in the stripping intensity
formula: $\Pi_{gal}$, $\rho_{0}$, $\sigma_{c}$ and ({\it ii}) the
shape of the two-dimensional stripping intensity function
$\eta$($\tilde{r}$,$\tilde{s}$) in the CPPS.


In the case of simulated clusters (top panel of figure
\ref{fig:stripp_eff_fig}) the region of lowest intensity (enclosed by
purple iso-contours) is a region of the CPPS where the non-accreted
galaxies are sheltered from ram-pressure gas stripping. In addition,
the large range of l-o-s velocities exhibited by the non-accreted
haloes in comparison with star-forming galaxies, could be
modulated/reduced by the ram-pressure effect as one can see from the
shapes of the contours of low stripping intensity $\eta{=}$0.1 and
0.3. In this respect, a population of stripped non-accreted galaxy
halos could be responsible for a decrease in the fraction of
non-accreted star-forming galaxy population with respect to the whole
non-accreted population. Furthermore, the fact that the galaxy
population in the field contains a important contribution of passive
galaxies \citep{Haines_et_al_2007} also reduces the fraction of the
non-accreted star-forming galaxy population with respect to the whole
non-accreted population. As the star-forming fraction and the
non-accreted fraction present similar values around 60-70\% in the
surroundings of virial regions, the two facts outlined above implies
that it would needed a population of accreted galaxies with a
remaining star formation currently located outside the virial radius.

Regarding the accreted halos, those star-forming galaxies which are
infalling for the first time through these CPPS regions of low
stripping intensity could survive as star-forming galaxies for a
fraction of the cluster crossing time (i.e. $\sim$1\,Gyr) before their
gas reservoir would be exhausted, see subsection
\ref{ssec:stfm_in_cpd} in this respect. Further, those accreted
galaxies which have never penetrated very deep in the virial region
with high velocities could survive for longer periods of time as
star-forming galaxies i.e. galaxies in circular orbits outside the
innermost parts of the virial region. In this respect, while early
spiral galaxies are in approximate isotropic orbits and the late
spiral and emission-line galaxies are clearly biased toward radial
orbits outside \mbox{$R_{P}{\sim}$0.7$r_{200}$}, there is a population
of galaxies in substructures which seems to be in circular orbits at
relatively large radii \citep{ENACS_XIII}.

As referred to the real clusters, the map of the stripping intensity
presents a region of very high intensity for those galaxies close to
the projected cluster center $\tilde{r}$$\sim$0 and with relatively
high velocities $\tilde{s}$$\gtrsim$0.5 (red iso-contours in
fig. \ref{fig:stripp_eff_fig}), this maximum of stripping intensity
can explain the paucity of star-forming galaxies in this specific
region of the CPPS. On the other hand, the region of very low
stripping intensity (purple iso-contours in
fig. \ref{fig:stripp_eff_fig}) presents a triangular shape, increasing
in separation towards larger radii and connected through a `bridge'
around ($\tilde{r}$,$\tilde{s}$)$\approx$(0,0). This allows to explain
the presence of star-forming galaxies near to the projected cluster
center, but only for those galaxies with relatively low l-o-s
velocities. In this respect, assuming that the radial profile of
projected-density along the transverse direction is similar to that
along the l-o-s direction, the geometrical effects introduce a minor
scatter (of the order of H$_{0}r_{200}$) in the l-o-s velocities
i.e. (H$_{0}r_{200}$/$\sigma_{c}$)$\sim$12\,\%. Thus, the observed
l-o-s velocity of galaxies near the projected center is a good proxy
for the actual l-o-s velocity without much information about its
position with respect to the cluster center. In consequence, those
star-forming galaxies with relatively low l-o-s velocities at the
projected center of clusters could come from any point along the
transverse clustercentric distance. However, assuming that they retain
a non negligible amount of star formation, it is quite likely that
they are well outside of the innermost part of the virial region.

As can be derived from the comparison of isocontours in the case of
real clusters (bottom panel of figure \ref{fig:stripp_eff_fig}), the
stripping intensity $\eta$($\tilde{r}$,$\tilde{s}$) and the CPPS density of
star-forming population are following opposite trends in the CPPS
where ({\it i}) the maximum of $\eta$ for $\tilde{r}$$\sim$0 and $\tilde{s}$${>}$0.5
coincides with a scarcity of star-forming galaxies, ({\it ii}) the
minimum for $\tilde{r}$${>}$1 and $\tilde{s}$${\lesssim}$0.5 coincides with the major
concentration of star forming galaxies and ({\it iii}) there is a
domain around ($\tilde{r}$,$\tilde{s}$)$\sim$(0,0) showing a low stripping intensity
and a significant density of star-forming galaxies. In particular, in
the low l-o-s velocity interval $\tilde{s}$${<}$0.5 of virial region, the
stripping intensity strongly modulates the CPPS distribution of
star-forming galaxies. This points out the map of stripping intensity
as a modulation function which is acting over the distribution
function in the CPPS.

As we have pointed out in the introduction, the effects of
environmental processes are cumulative. Some of them can operate
simultaneously quenching star formation in galaxies and these
environmental processes operate with different time-scales, regions of
influence and varying their intensity depending on galaxy properties
e.g. galaxy gravitational mass. This is one of the hypothesis with
which we work in \citet{Hernandez-Fernandez_et_al_2012b}. Furthermore,
only the appropriate observable can trace the effect of a specific
environmental process. In this work, we probe the CPPS distribution of
star-forming population and this space includes the proxies of the two
physical variables which mainly determine the intensity of
ram-pressure stripping i.e. cluster-centric radius and galaxy
velocity. Therefore, the opposite trend in the CPPS between the
density of star-forming population and the intensity of stripping
clearly points out that the effect of ram-pressure stripping is a
major contributor to quench the star formation in cluster galaxies in
virial regions and in their close surrounding regions.

\subsection{Comparison between the CPPS distributions of intense and quiescent star-forming galaxies.}
\label{ssec:stfm_in_cpd}

\begin{figure}[!ht]
\centering
\resizebox{1.00\hsize}{!}{\includegraphics{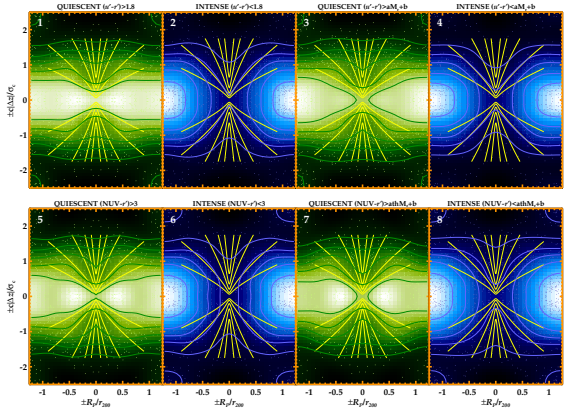}}
\caption{\bf Stacked CPPS diagrams for the star-forming population
  separated between the intense star-forming galaxies and the
  quiescent star-forming galaxies. From top to bottom and left to
  right, odd (even) panels represent quiescent (intense) star-forming
  galaxies split them by ({\it a}), ({\it b}), ({\it c}) and ({\it d})
  conditions as it is described in subsection
  \ref{ssec:stfm_in_cpd}. Isodensity lines and the rest of elements of
  the figure are defined in the same way as figure
  \ref{fig:high-lum_gal1} and contours of iso-$\eta$ have the same
  values that figure \ref{fig:stripp_eff_fig}.}
\label{fig:stfr_quiescent-vs-intense2}
\end{figure}

In this subsection, we probe the distribution of star-forming galaxies
in the CPPS depending on the intensity of their star formation. A
number of tests have been performed on the criterion to segregate
between intense and quiescent star-forming galaxies in the two
color-magnitude diagrams \mbox{(NUV$-r'$ vs. $M_{\it r'}$)} and
\mbox{($u'{-}r'$ vs. $M_{\it r'}$)}. These tests include a segregation
based on fixed-color cuts and also tilted/curved lines in
color-magnitude diagrams to account for the trend of `blue cloud'
towards bluer colors with decreasing luminosity
\citep{Wyder_et_al_2007}. The criterion applied $u'{-}r'{=}$1.8 seems
to be the one producing a clearer segregation between intense and
quiescent star-forming galaxies in the CPPS. The result of the
different criterion are shown in figure
\ref{fig:stfr_quiescent-vs-intense2}. From top to bottom and left to
right, the criterion to segregate intense from quiescent star-forming
galaxies are: ({\it a}) $u'{-}r'{=}$1.8 attempts to split the
star-forming population in two samples of similar size, ({\it b})
$u'{-}r'{=}$1.75${-}$0.20($M_{\it\,r'}{+}$20) traces the peak of the
blue sequence along the observed range of $r'$-band luminosity, ({\it
  c}) NUV${-}r'{=}$3 separates the green valley from blue sequence
with a unique color value, ({\it d}) NUV${-}r'{=}$3.590
${+}$0.075($M_{\it\,r'}$+20) ${-}$0.808\,{\it tanh}(($M_{\it
  r'}$+20.32)/1.81) separates the green valley from the blue sequence
taking as reference the fit to the blue sequence provided by
\citet{Wyder_et_al_2007}.


The distribution of star-forming galaxies in the CPPS depending on
their star-formation intensity shows two evident trends: ({\it 1}) the
intense star-forming galaxies are preserved outside the projected
virial radius in the outer parts of the region of lowest stripping
intensity and ({\it 2}) the quiescent star-forming galaxies are
distributed along the whole observed radial range and partially
concentrated around two maxima, one maximum is located inside the
projected virial radius but apart from the cluster center and the
other maximum appears outside the projected virial radius. The first
point suggests that the intense star-forming galaxies are those field
galaxies with higher levels of star formation which have not yet
suffered the star-formation quenching by the environment of cluster
virial regions while the quiescent star-forming galaxies are shared
between those field galaxies with lower levels of star formation and
those star-forming galaxies showing a decline in their star formation
due to the environmental effects of cluster virial regions. Accepting
this description, in the case that the bulk star-forming galaxies were
galaxy interlopers, an important fraction of intense star-forming
galaxies would be observed inside the projected virial radius. In
addition, the density of quiescent star-forming galaxies around
($\tilde{r}$,$\tilde{s}$)$\sim$(0,0) shows a clear impact of the
stripping phenomenon as is suggested by the similar trends of both the
isodensity contours of quiescent galaxies and the iso-intensity
contours of gas stripping in this region of the CPPS. These results
reinforce the hypothesis that ram-pressure stripping is strongly
influencing in the modulation of star formation history of cluster
galaxies and specifically, of the star forming galaxies. We stress
that this clear segregation in the CPPS between intense and quiescent
star-forming galaxies is found applying different criteria about two
colors, NUV${-}r'$ and $u'{-}r'$. In contrast, a so clear segregation
is not found when the same sample of star-forming galaxies is split by
their $r'$-band luminosity between high-luminosity
($M_{\it\,r'}{<}{-}$19.5) and low-luminosity galaxies
($M_{\it\,r'}{>}{-}$19.5), a galaxy property which strongly correlates
with galaxy mass. Then, it is not expected that this segregation in
the CPPS between intense and quiescent star-forming galaxies would
come from differences in their stellar mass distributions. In this
respect, \citet{ENACS_XI} find that the different morphological types
(E, S0, early and late spiral and emission-line galaxies) present
different CPPS distributions but it is only found an evident
luminosity segregation for the brightest $M_{R}{<}{-}$22 ellipticals
that are not in galaxy substructures.

In the top left panel of figure \ref{fig:stfr_quiescent-vs-intense2},
it can be seen that the intense star-forming galaxies $u'{-}r'{<}$1.8
are mainly populating the space just outside the projected virial
radius while the quiescent star-forming population $u'{-}r'{>}$1.8 is
formed by those galaxies outside the projected virial radius and more
remarkable, a population of galaxies clearly biased towards the
center. We try to explain the most clear segregation between quiescent
and intense star-forming galaxies splitting them at $u'{-}r'$=1.8,
invoking a `synchronisation' between two natural time-scales: the
timescale of variation of visible color(s) \citep{Kennicutt_1998} and
the crossing time for clusters \citep{Boselli&Gavazzi_2006} which are
around 10$^{9}$ yr. This coincidence has the consequence that an
infalling star-forming galaxy shows, by means of its visible color
$u'{-}r'$, that its star-formation activity is drastically diminished
just when it would be well inside the cluster. This is only applied to
those star-forming galaxies modestly attenuated, given that those
star-forming galaxies highly attenuated appear in the red sequence
\citep{Wolf_et_al_2009b}.

\section{Summary and conclusions}
\label{sec:summ&conclu}

We summarize the main results and conclusions of this work in the
following paragraphs:

\begin{itemize}

\item There is no statistically significant luminosity segregation in
  the CPPS distribution of star-forming and passive galaxy populations
  along the observed luminosity range
  $-$23${\lesssim}M_{\it\,r'}{\lesssim}{-}$18.

\item The passive population is concentrated around a unique maximum
  in the CPPS clearly centered at ($\tilde{r}$,$\tilde{s}$)$\sim$(0,0)
  as it was found by previous works in the literature. The `boxy"
  and/or `peanut' shape of isodensity contours of passive population
  seems to be a direct consequence of a projected density radial
  profile varying with ${\sim}\tilde{r}^{-1}$ in the innermost parts
  of the virial region. This radial trend can be mainly obtained from
  a more cuspy profile where the uncertainties in cluster centering
  are included.

\item The star-forming population presents a maximum around
  ($\tilde{r}{\sim}$0,$\tilde{s}{\gtrsim}$1) showing an evident
  preference for environments outside the virial region and with lower
  l-o-s velocities. The significant fraction of star-forming galaxies
  at the projected center of clusters are mainly those galaxies with
  low l-o-s velocities and they can be mainly identified as those
  galaxies with a remaining star formation activity inside the
  physical virial region or, in a lower degree, as galaxy interlopers
  i.e. outside the physical virial region. The CPPS density of
  star-forming galaxies and the intensity of ram-pressure stripping
  $\eta$($\tilde{r}$,$\tilde{s}$) present an opposite trend throughout
  the CPPS. This opposite behaviour suggests there is a major
  contribution of ram-pressure stripping to the modulation of
  star-formation activity of galaxies in cluster virial regions and
  their close vicinity.

\item The intense star-forming population is mainly located around an
  unique maximum outside the virial region and around
  ($\tilde{r}$,$\tilde{s}$)$\sim$(0,1.25). This regions of the CPPS
  coincides with the outer part of the regions with lowest stripping
  intensity. The quiescent star-forming population is distributed in a
  more extended range of radii, also occupying the CPPS region of the
  lowest stripping intensity but as much inside the projected virial
  radius as outside the projected virial radius.

\item In the cross-identification of the passive population with the
  accreted population and also, of the star-forming population with
  the non-accreted population, we find that:

\begin{itemize}

\item the non-accreted population seems to make a major contribution
  to the star-forming population outside the virial region where the
  larger range in l-o-s velocity shown by the non-accreted population
  in comparison with the star-forming population can be reconciled
  taking into account the effect of the ram-pressure stripping. It
  would be needed a population of accreted galaxies with a remaining
  star formation inside the virial radius $R_{P}{<}r_{200}$ to reach
  the fraction of star-forming galaxies inside the virial
  radius. Considering the similar fraction of the non-accreted
  population and the star-forming population outside the virial
  radius, it also would be needed a population of accreted galaxies
  with a remaining star formation outside the virial radius
  $R_{P}{>}r_{200}$ in order to counteract a population of stripped
  non-accreted galaxies and the population of passive galaxies from
  the field i.e. non-accreted passive galaxies.

\item the inner logarithmic slope of the radial profile of projected
  density of accreted population $\alpha{\sim}-$1.19 can be
  substantially reduced when are considered the uncertainties of
  cluster centering. This decrease in the inner slope can be enough to
  reach the value of the inner slope of the density profile of the
  passive population $\alpha{\sim}-$1. The velocity distribution of
  accreted and passive populations are similar within
  uncertainties. We can conclude that the accreted population can
  account for the bulk of the passive population at the innermost part
  of virial region.

\item the star-forming population of the inner parts of virial regions
  appears to mainly come from that accreted population with a
  relatively low l-o-s velocity and a remaining star formation
  activity and also, come from the non-accreted population
  $r{>}r_{200}$ which appears projected inside the projected virial
  radius $R_{P}{<}r_{200}$ i.e. they are galaxy interlopers.

\end{itemize}

\end{itemize}

\section*{Acknowledgements}
\label{sec:acknowledgements}

J.D.H.F. acknowledges support through the FAPESP grant project
2012/13381-0. C.P.H. was funded by CONICYT Anillo project
ACT-1122. A.D. acknowledges partial support from the INFN grant Indark
and from the grant Progetti di Ateneo/CSP TO\_Call\_2\_2012\_0011
`Marco Polo' of the University of Torino. J.I.P. acknowledges support
through Proyecto de Excelencia FQM7058 ``Historia de formaci{\'o}n
estelar y evoluci{\'o}n qu{\'i}mica de galaxias en entornos de
diferente densidad". C.M.d.O. acknowledges support through FAPESP
project 2006/56213-9 and CNPq grant 305205/2010-2. J.M.V. acknowledges
support from the project AYA2010-21887-C04-01 ``Starbursts and their
imprint in the cosmic evolution of galaxies".

The formal acknowledgements to the resources used in this work can be
read at:
www.sdss.org\allowbreak{}/dr6\allowbreak{}/coverage\allowbreak{}/credits.html
for the Sloan Survey, ned.ipac.caltech.edu for the NED webpage and
galex.stsci.edu\allowbreak{}/GR6\allowbreak{}/?page=acknowledgments
for the GALEX mission.


\section*{Appendix}
\label{sec:appendix}

It is widely known the handicap of the fiber collision in the Main
Galaxy Sample of SDSS; two spectroscopic fibers cannot be placed
closer than 55 arcsec on a given plate
\citep{Strauss_et_al_2002}. This produces a lack of completeness
throughout the survey which could be dramatic in the case of highly
crowded fields e.g. the central regions of galaxy
clusters. \citet{Park&Hwang_2009} computed the radial dependence (in
angular units) of the completeness of the Main Galaxy Sample for a
sample of Abell clusters. They found a radial trend which goes from
the asymptotic value around 90\,\% well out of cluster regions up to a
completeness in the interval of 75-80\,\% in the very center of
clusters. Applying this completeness correction in galaxy clusters
along the radial axis has the disadvantage of including, in both the
numerator and the denominator of the completeness fraction, a galaxy
population which is not a genuine cluster population. This can bias
the estimation of the completeness for a genuine cluster galaxy
sample.

\begin{figure}[!ht]
\centering
\resizebox{1.00\hsize}{!}{\includegraphics{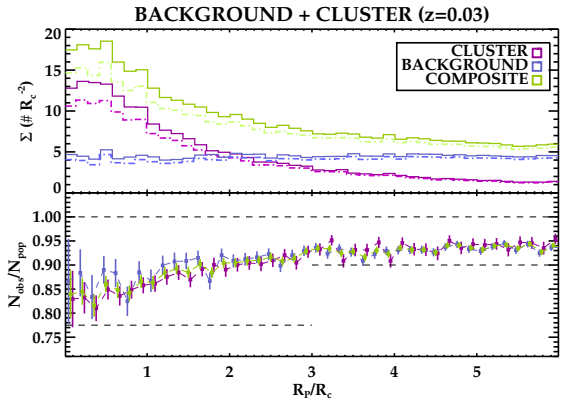}}
\caption{\bf Simulated completeness of the Main Galaxy Sample
  depending on clustercentric radius in core radius unit $R_{c}$. Top
  graph: Radial profiles of projected density from each population;
  cluster (purple), background (blue) and composite (green). The dark
  solid histograms correspond to the whole population and the light
  dashed histograms are the non-rejected population. Bottom graph:
  Completeness of each population with respect to its own total
  population. The lines including binomial uncertainty bars are the
  fractions of the non-rejected population over its own total
  population. Dashed flat lines mark the 90 and 77.5\,\% which are the
  tendencies for small and large radii found in
  \citet{Hwang&Lee_2008}. Also, it is shown a dashed line at 100\,\%
  for comparison purposes.}
\label{fig:rad_comp}
\end{figure}

In order to estimate this fraction in an unbiased way and test the
trend of this fraction along the radial axis, we simulate a composite
population of a massive cluster at redshift $z$=0.033 embedded into a
galaxy background population. We make the test at this redshift
because most clusters in this work are around this redshift. We assume
that the galaxy projected density of a cluster follows a generalized
King profile with the physical parameters found by
\citet{Abdullah_et_al_2011}: $\gamma{=}{-}$(2/3) and a median value
for the core radius of \mbox{$R_{c}$=0.295\,$h^{-1}$\,Mpc}. So, the
angular distance of 55 arcsec in unit of this core radius for a
cluster at $z$=0.033 is d(55")=0.085$R_{c}$. We apply an algorithm of
rejection which mimics the algorithm proposed by
\citet{Tiling_algorithm_2003}, rejecting with higher priority those
objects with a larger set of neighbour distances smaller than the
critical distance d=55". We tune the projected density of background
objects to reproduce at large radii the average projected density of
objects of the Main Galaxy Sample \mbox{$\Sigma{\sim}$190 gal
  deg$^{-2}$} which correspond to a projected density of
$\Sigma{\sim}$6.5\,gal\,$R_{c}^{-2}$ in natural units. The richness of
the cluster is fixed to $\sim$250 galaxies inside the $r_{200}$ taking
as reference ABELL\,1185 from the set of clusters under study in this
work. The stacked result from 50 trials is summarized in figure
\ref{fig:rad_comp}.


In a first inspection, the completeness of the composite
(cluster+background) population follows a similar trend to the one
found by \citet{Park&Hwang_2009} which has a mean value of 75-80\,\%
at the very center of the cluster, whereas it shows an asymptote
around 90\,\% at large radii. In this simulated trial, the trend of
clustercentric completeness shows a bias of around 5 per cent toward
higher values with respect to the one found by
\citet{Park&Hwang_2009}. Given that we deal with a background
population of uniform density without any contribution of galaxy
substructures, we attribute the lower completeness in the real sample
to the presence of galaxy pairs and groups. This is not just easily
applied to the field well outside the clusters, where there is a
significant contribution of galaxy pairs and groups to the whole
galaxy population \citep{Cox_2000}, but also at the very cluster
center, where it is found a small contribution of galaxy substructures
to the cluster population \citep[][]{ENACS_XI}.

The completeness in the three populations shows a monotonic decreasing
trend with projected density as it would be expected. Specifically in
the case of the composite population, the completeness decreases from
93-94\,\% for a density of $\Sigma{\sim}$6.5\,gal\,$R_{c}^{-2}$ to a
completeness of around 85\,\%, this is a reduction of around a 10\,\%,
for a density approximately three times higher,
$\Sigma{\sim}$18\,gal\,$R_{c}^{-2}$. Along the radius range under
$R_{c}$=2, the cluster completeness is not only systematically lower
than the completeness from the composite population but also the
completeness of the background population. It would be explained
assuming a composite population where two populations with two
different projected densities are mixed together: (1) as the projected
density increases, the number of objects with neighbour distances
smaller than the critical distance increases, (2) in a composite
population, a new and different set of links smaller than the critical
distance between objects from the two populations is established, (3)
in these newly established links, the algorithm will penalize with
higher priority the objects from a population with a higher density
than the objects from a population with a lower projected density.

From this simple test, we can conclude that a genuine cluster
population on top of a background population does not show a very
different completeness from the composite population, assuming the
range of galaxy projected densities similar to those we are working
with. Indeed, for higher redshifts the apparent size of a cluster will
be reduced and in consequence, the galaxy projected densities should
be large for a cluster galaxy sample with an equivalent completeness
limit in luminosity. In addition, the projected density of the
background population also increases for the fainter completeness
limits which are required for observations at higher redshifts. These
two facts contribute to reduce the completeness of the non-rejected
populations.


\bibliographystyle{apj} 
 \bibliography{bibliografia}

\end{document}